\documentclass[a4paper,fleqn,usenatbib]{mnras}

\usepackage{newtxtext,newtxmath}
\usepackage[T1]{fontenc}
\usepackage{ae,aecompl}

\usepackage{graphicx}	% Including figure files
\usepackage{amsmath}	% Advanced maths commands
\usepackage{amssymb}	% Extra maths symbols
\usepackage{CJKutf8}
\usepackage{multirow}

\newcommand{\kilo}{{\rm k}}
\newcommand{\msun}{M_\odot}
\newcommand{\mjup}{M_{\rm J}}
\newcommand{\vinf}{V_{\rm inf}}

\newcommand{\nbodyplus}{{\tt NBODY6++}}
\newcommand{\nbodylong}{{\tt NBODY6++GPU}}

\newcommand{\rebound}{{\tt rebound}}
\newcommand{\amuse}{{\tt AMUSE}}
\newcommand{\hdf}{{\tt HDF5}}

\def\apgt{\ {\raise-.5ex\hbox{$\buildrel>\over\sim$}}\ }
\def\aplt{\ {\raise-.5ex\hbox{$\buildrel<\over\sim$}}\ }
\def\lteq{\ {\raise-.5ex\hbox{$\buildrel<\over-$}}\ }

\title[Multiplanetary Systems in Star Clusters]{Stability of Multiplanetary Systems in Star Clusters}

\author[M. X. Cai et al.]{
Maxwell Xu Cai (\begin{CJK*}{UTF8}{gbsn}蔡栩\end{CJK*})$^{1,3,4}$\thanks{E-mail: cai@strw.leidenuniv.nl (MXC)},
M.B.N. Kouwenhoven$^{2,4}$,
Simon F. Portegies Zwart$^{1}$,
\newauthor and Rainer Spurzem$^{3,4,5}$
\\
$^{1}$Leiden Observatory, Leiden University, PO Box 9513, 2300 RA, Leiden, The Netherlands\\
$^{2}$Department of Mathematical Sciences, Xi'an Jiaotong-Liverpool University (XJTLU), 111 Ren'ai Road, Dushu Lake Higher Education\\ Town, Suzhou Industrial Park, Suzhou 215123, P.R. China\\
$^{3}$National Astronomical Observatories and Key Laboratory of Computational Astrophysics, Chinese Academy of Sciences, 20A Datun Road,\\ Chaoyang District, Beijing 100012, P.R. China\\
$^{4}$Kavli Institute for Astronomy and Astrophysics, Peking University, 5 Yi He Yuan Road, Haidian District, Beijing 100871, P.R. China\\
$^{5}$Astronomisches Rechen-Institut, Zentrum f\"ur Astronomie, University of Heidelberg, M\"onchhofstrasse 12-14, 69120 Heidelberg, Germany
}

\date{Accepted 2017 June 9. Received 2017 June 7; in original form 2016 September 27}

\pubyear{2017}

\begin{document}
\label{firstpage}
\pagerange{\pageref{firstpage}--\pageref{lastpage}}
\maketitle

\begin{abstract}
Most stars form in star clusters and stellar associated. However, only about $\sim 1\%$ of the presently known exoplanets are found in these environments. To understand the roles of star cluster environments in shaping the dynamical evolution of planetary systems, we carry out direct $N$-body simulations of four planetary systems models in three different star cluster environments with respectively $N=2\kilo, 8\kilo$ and $32\kilo$ stars. In each cluster, an ensemble of initially identical planetary systems are assigned to solar-type stars with $\sim 1 \msun$ and evolved for 50~Myr. We found that following the depletion of protoplanetary disks, external perturbations and planet-planet interactions are two driving mechanisms responsible for the destabilization of planetary systems. The planet survival rate varies from $\sim 95\%$ in the $N=2\kilo$ cluster to $\sim 60\%$ in the $N=32\kilo$ cluster, which suggests that most planetary systems can indeed survive in low-mass clusters, except in the central regions. We also find that planet ejections through stellar encounters are cumulative processes, as only $\sim 3\%$ of encounters are strong enough to excite the eccentricity by $\Delta e \geq 0.5$. Short-period planets can be perturbed through orbit crossings with long-period planets. When taking into account planet-planet interactions, the planet ejection rate nearly doubles, and therefore multiplicity contributes to the vulnerability of planetary systems. In each ensemble, $\sim 0.2\%$ of planetary orbits become retrograde due to random directions of stellar encounters. Our results predict that young low-mass star clusters are promising sites for next-generation planet surveys, yet low planet detection rates are expected in dense globular clusters such as 47 Tuc. Nevertheless, planets in denser stellar environments are likely to have shorter orbital periods, which enhances their detectability.

\end{abstract}


%%%%%%%%%%%%%%%%%%%%%%%%%%%%%%%%%%%%%%%%%%%%%%%%%%

%%%%%%%%%%%%%%%%% BODY OF PAPER %%%%%%%%%%%%%%%%%%

\section{Introduction}

Research of planetary systems can be dated back to at least a few thousand years ago, when astronomers began to observe the night sky and named the planets in the Solar System ``wanderers''. Observations of star clusters started several centuries ago, after telescopes became powerful enough to resolve these ``clouds of stars''. Only in the recent decades the possible relationship between planetary systems and star clusters was gradually recognised. Thanks to the availability of dedicated observational facilities such as Kepler, more than 3\,600 extrasolar planetary systems have now been identified, among which about 610 are multiplanetary systems\footnote{\href{http://exoplanet.eu}{http://exoplanet.eu}}. Notably, approximately 20 exoplanets have been detected in star clusters (see Table~\ref{tab:p_sys_sc}). The discoveries of exoplanets in star clusters show that star clusters contain a variety of celestial bodies, and therefore it is important to further study these star clusters and investigate the planetary systems that they host. 

\begin{table*}
	\centering
	\caption{List of exoplanet detections in star clusters. DM: detection method; TS: transit; RV: radial velocity; TM: timing; Nep: Neptune-sized; $M_{\rm S}$: Stellar mass in solar units $\msun$; $m_{\rm p}$: planet mass in Jupiter units $M_{\rm J}$; $P$: orbital period in days.}
	\begin{tabular}{lrrrlll}
		\hline
		\hline
		{\bf Designation} & {\bf $m_{\rm p}$ ($M_{\rm J}$)} & {\bf $P$ (days)} & {\bf $M_{\rm S}$ ($\msun$)} & {\bf DM} &  {\bf Cluster} & {\bf Reference} \\
		\hline
		\hline
		YBP401 b          & 0.46   & 4.09   & 1.14  & RV & M67     & \cite{Brucalassi16} \\	
		Pr0211c           & 7.9    & 5\,300 & 0.935 & RV & M44     & \cite{malavolta2016} \\
		EPIC-210490365 b  & $< 3$  & 3.49   & 0.29  & TS  & Hyades   & \cite{mann2016}  \\
		SAND 364 b        & 1.5    & 121.7  & 1.35  & RV & M67     & \cite{Brucalassi14}  \\
		YBP1194 b         & 0.34   & 6.96   & 1.01  & RV & M67      & \cite{Brucalassi14} \\
		YBP1514 b         & 0.4    & 5.11   & 0.96  & RV & M67      & \cite{Brucalassi14} \\
		HD 285507 b       & 0.92   & 6.08   & 0.73  & RV & Hyades   & \cite{quinn2014} \\
		Kepler-66 b       & Nep    & 17.82  & 1.04  & TS  & NGC6811  & \cite{meibom13} \\
		Kepler-67 b       & Nep    & 15.73  & 0.87  & TS  & NGC6811  & \cite{meibom13} \\
		Pr0201b           & 0.54   & 4.33   & 1.234 & RV & M44     & \cite{quinn2012} \\
		Pr0211b           & 1.88   & 2.15   & 0.935 & RV & M44     & \cite{quinn2012}  \\
		eps Tau b         & 7.6    & 595    & 2.70  & RV & Hyades   & \cite{sato2007}  \\
		NGC 2423 3 b      & 10.6   & 714    & 2.40  & RV & NGC2423  & \cite{lovis2007} \\
		NGC 4349 No 127 b & 19.8   & 678    & 3.90  & RV & NGC4349  & \cite{lovis2007} \\
		PSR B1620-26 b    & 2.50   & 36\,525 & 1.35  & TM & M4      & \cite{backer1993} \\
		\hline
		\hline
	\end{tabular}
	\label{tab:p_sys_sc}
\end{table*}

Modern planet formation theories, such as the core accretion scenario \citep[e.g.,][]{mizuno1980, bodenheimer1986} and the disk-fragmentation scenario \citep[e.g.,][]{boss1997}, were inspired by the Nebular Hypothesis proposed independently by Immanuel Kant and Pierre-Simon Laplace. According to these theories, planets form in the protoplanetary disks, are essentially byproducts of the star formation process. Consequently, planets are may be common around main sequence stars. The recent discoveries of many exoplanets indeed have confirmed the high frequencies of exoplanets. 

Moreover, observational studies suggest that most stars form in groups and/or clusters following the collapse of giant molecular clouds \citep[e.g.,][]{clarke2000, lada2003, porras2003}. Naturally, one would expect that planets also form in star clusters. Indeed, isotope analysis of meteorites suggests that even our own Solar system may have formed in a star cluster \citep[see, e.g.,][and references therein]{adams2001, portegies09l, Pfalzner2013, parker2014, Pfalzner2015}. Direct imaging surveys have revealed protoplanetary disks in the Trapezium cluster, which is embedded in the Orion Nebula \citep{McCaughrean1996,bally2000}. These disks are likely the precursors of planetary systems. The orbits of outer solar system bodies such as Sedna may also be explained as a consequence of close encounters, \citep{adams10,jilkova2015}. If planetary systems are formed in star clusters, their dynamical evolution will then be influenced not only by internal processes \citep[e.g.][]{voyatzis2013, zhang2013, liy2015,liy2016} but also by the dynamical evolution of their host star clusters. The orbital architectures of the planetary systems discovered in the Galactic field can be thus used to constrain the properties of the star clusters in which they were formed.

As of May 2017, only 1\% of the known exoplanets are in star clusters (as listed in Table~\ref{tab:p_sys_sc}). This apparent low frequency of exoplanets in star clusters can partially be attributed to observational selection effects. Nevertheless, theoretical studies indicate that external perturbations due to stellar flybys play a role in the evolution of planetary systems \citep[e.g.,][]{heggie96, laughlin98, davies01,bonnell2001}. Naturally, an immediate question arises on how the dynamical evolution of planetary systems depends on the hosting stellar environments. Galaxies and star clusters are distinctively different stellar environments, in the sense that the they have very different relaxation timescales \citep{binney2008}. The relaxation time in a typical galaxy of $N \sim 10^{11}$ stars is well over a Hubble time, and therefore close encounters are unimportant (except for the galactic center); exoplanets around stars in the galactic field rarely experience external perturbation due to stellar encounters. In contrast, the corresponding relaxation time in open clusters is of the order of $\sim$~Myr. Should there be any planetary systems in star clusters, they are likely to experience external perturbations. This motivates \cite{portegies15} to define the ``parking zone'' of planetary systems: a range of orbital separations around any star within which the planets native to that star may have been perturbed by close encounters of external perturbers, but are unlikely to be affected by external perturbations once the star becomes a member of galactic field population. A number of recent numerical studies have shown that frequent close encounters that can be destructive for planetary systems \citep[e.g.,][]{spurzem09, malmberg2011, hao2013, liu2013, li2015, zheng2015, wang2015b, shara2016}. Therefore, it would be insightful to carry out an exploratory study to understand the role of dynamical stellar environment on the evolution of planetary systems.

Due to the chaotic nature of $N$-body systems with $N \ge 3$, the research on analytical formulations of generic multiplanetary systems remains extremely challenging. We therefore carry out the investigation with gravitational direct $N$-body simulations. However, given the enormous spatial and temporal differences between the dynamics of star clusters and planetary systems, numerical investigation of planetary systems in star clusters is non-trivial, as it leads to a highly hierarchical and chaotic many-body systems. Substantial progress was made over the years with studies of single-planet systems in star clusters. Using direct $N$-body simulations, \cite{hurley02} investigate the fate of free-floating planets, and find that planets in star clusters, after being liberated by stellar encounters, can remain bound in open clusters for a half-mass relaxation time scale of the star cluster. In another direct $N$-body study, \cite{spurzem09} investigate the dynamics of single-planetary systems in star clusters using both direct $N$-body simulations and hybrid Monte-Carlo simulations. Their results indicate that compact and nearly-circular orbits are generally not affected by distant stellar encounters. On the other hand, while short-period planets are difficult to perturb, close encounters can excite these to moderate eccentricities, which may in turn result to orbital decay due to tidal dissipation. \cite{zheng2015} discuss the effect of the initial virial state and the presence of initial substructure, and find that single-planet systems wider than approximately 200~AU are mostly vulnerable by the time the cluster reaches an age of 50~Myr.

Studies of multiplanetary systems in star clusters have thus far been limited to Monte Carlo scattering experiments. \cite{hao2013} conclude that multiplanetary systems are affected by both direct interaction with the encountering star and planet-planet scattering. The combined effects can account for the apparent low frequency of exoplanets in star clusters, not only for those on wide orbits that are directly affected by stellar encounters, but also planets close to the star which can disappear long after a stellar encounter has perturbed the outskirts of the planetary system. A more recent study by \cite{li2015} extends the simulations to a much wider parameter space with over two million scattering simulations. By varying the compactness of the target solar systems, the velocity dispersion of the host star cluster, stellar host masses, starting eccentricities of planet orbits, and single versus binary perturbers, they characterised the encounter cross sections as a function of stellar host mass, cluster velocity dispersion, semi-major axis, and final eccentricity, and predicted that the Solar System was formed within an open cluster with $N \lesssim 10^4$ stars.

In the case of stars with two planetary companions, it is possible to employ a variety of three-body approaches. \cite{mardling08} proposes an analytical criterion for determining the stability of arbitrary three-body hierarchies which makes use of the chaos theory concept of resonance overlap. In a follow-up paper, the three-body stability algorithm given in \cite{mardling08} is used to determine the stability of an ensemble of mini solar systems with two Jupiter-mass planets in open cluster environments \citep{shara2016}. 

Star cluster environments not only affect the post-formation dynamical evolution of planetary systems, but also affect the planet formation process through the circumstellar disks \citep[e.g.,][]{thies2010, thies2011, anderson13}. \cite{adams04} points out that the intensive radiation from nearby OB stars can modify the mass-loss rate and evaporation timescales of exposed circumstellar disks, and eventually affect the planet-formation processes and and planet migration through disk-planet interaction. \cite{olczak2012} carries out simulations to study the mass-loss process driven by intensive radiation at the Arches cluster environment. They find that stellar encounters destroy one-third of the circumstellar disks in the cluster core within the first 2.5~Myr of evolution, and after 6 Myr half of the core population becomes diskless. However they also point out that the disk destruction process ceases after roughly 1~Myr in sparser clusters due to significant cluster expansion \citep{olczak2010}. A recent study by \cite{portegies2016} shows that close encounters result in the truncation of circumstellar disks. This mechanism can be used to reproduce the observed  distribution of disk sizes in the Orion Trapezium cluster. Furthermore, he shows that a subvirial ($Q \sim 0.3$) and fractal ($F \sim 1.6$) initial environment is indicated according to the observed disk size distribution.

In most previous numerical experiments, the star cluster dynamics and planetary system dynamics are integrated by a single code. For example, \cite{spurzem09} carried out direct $N$-body simulation with the code \nbodyplus{} \citep{spurzem99} by initializing single planets as KS-binary with their host stars. This is highly accurate thanks to the KS regularization \citep{ksreg65} technique, yet this is prohibitively expensive, even for moderately large systems, and inefficient in handling multiplanetary systems. The Monto-Carlo approach is computationally quite affordable, but the results depend on the assumed distribution of the perturber's velocity and impact parameter.
Finally, free-floating planets (FFP) are expected in star clusters following the ejections from the host stars \citep[see, e.g.,][and references therein]{kouwenhoven2016}. 
%This study aims to provide insights into the FFP population, which is currently observed mainly through the microlensing technique.

The host star clusters themselves are also evolving, driven by internal mechanisms such as two-body relaxation \citep[e.g.,][]{Chandrasekhar42, spitzer71,henon71,takahashi00} and stellar evolution \citep[e.g.,][]{applegate86, portegies98, kouwenhoven2014}, as well as external mechanisms such as the interaction with the Galactic tidal field \citep[e.g.,][]{spitzer87,cai2016} and the spiral arms (e.g., \citealt{gieles07}). In order to take all these effects into account, we carry out direct $N$-body simulations with \nbodylong{} \citep{wang2015a, wang2016, aarseth03, spurzem99}, and obtain the properties of all stellar encounters in these simulations. Subsequently, the perturbation data are loaded into the {\tt AMUSE} framework \citep[e.g.,][]{portegies09, portegies13, mcmillan2012, Pelupessy13} and are sent to the planetary systems integrator \rebound{} \citep{rein12}, where they are included in the modelling of the planetary systems.

This paper is organized as follows. The modeling of the perturbations and the implementation of the simulations are presented in Section~\ref{sec:modeling}. The initial conditions of the star clusters and planetary systems are described in Section~\ref{sec:ic}. The results are presented and discussed in Section~\ref{sec:results}. Finally, the conclusions are presented in Section~\ref{sec:conclusions}.

%%%%%%%%%%%%%%%%%%%%%%%%%%%%%%%%%%%%%%%%%%%%%%%%%%%%%%%
%%%%%%%%%%%%%%%%%%%%%%%%%%%%%%%%%%%%%%%%%%%%%%%%%%%%%%%
%%%%%%%%%%%%%%%%%%%%%%%%%%%%%%%%%%%%%%%%%%%%%%%%%%%%%%%
\section{Modeling and Implementation} \label{sec:modeling}
\subsection{Modeling Perturbations} 

\begin{figure}
\begin{center}
	\includegraphics[scale=0.35]{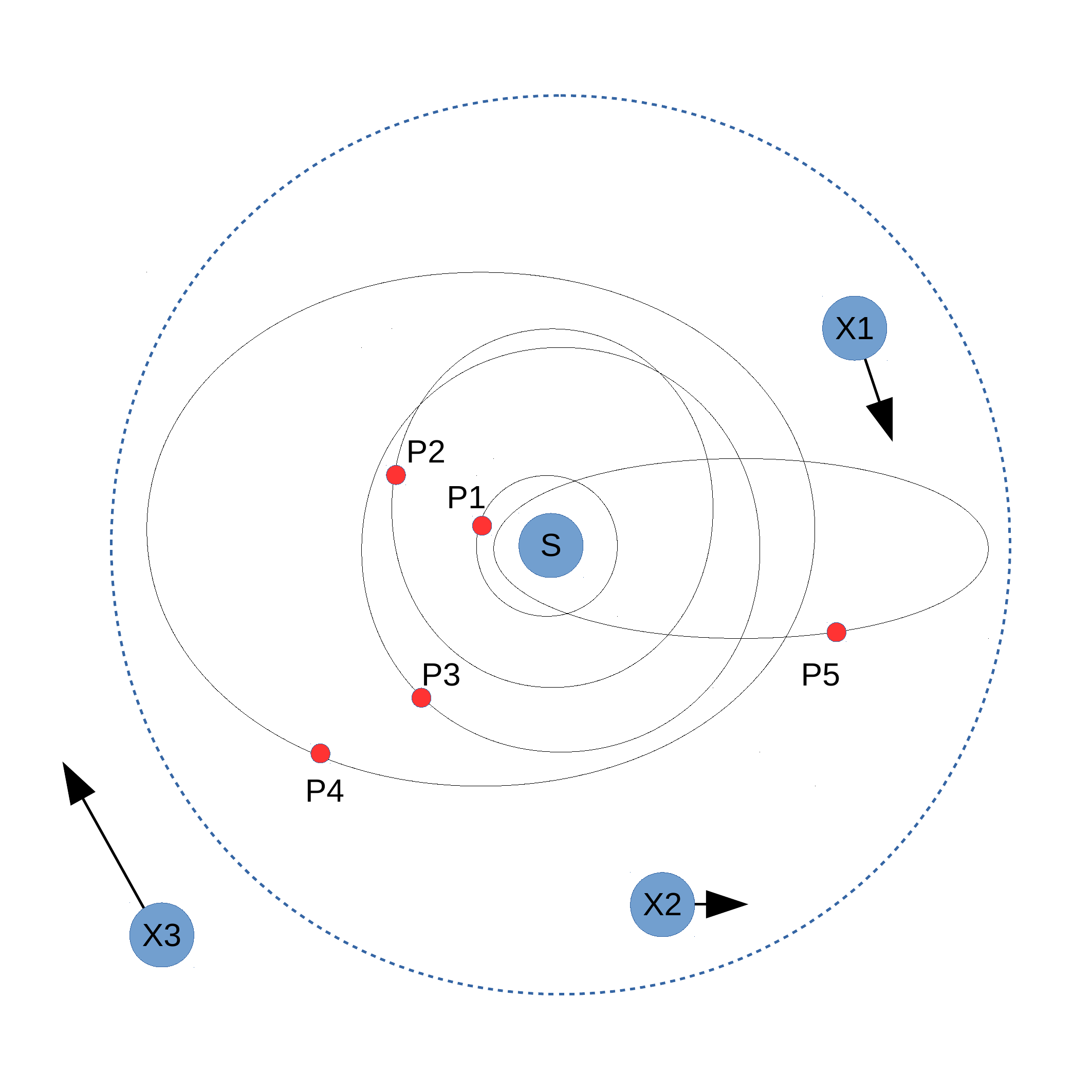}
\end{center}
\caption{Schematic illustration of a planetary system consisting of a host star (S) and five planets (P1 to P5), being perturbed by three external perturbers (X1, X2, X3). The dotted circle indicates the boundary of the neighbor sphere, beyond which we ignore the influence of any perturbations (distance not to scale). The arrows indicate the directions and magnitudes of the perturbers' velocities. In this example, the perturbations due to X1 and X2 are taken into account, whereas the perturbation due to X3 is neglected. Slow perturbers are more likely to remain in the neighbor sphere for longer time, causing stronger perturbations. Note that when a planet's orbit is sufficiently excited, its orbit may intersect with the orbits of inner planets.}
\label{fig:pert_schematic}
\end{figure}

It is convenient to work in a Cartesian coordinate system centered at the host star of the planetary system under consideration, as illustrated in Figure~\ref{fig:pert_schematic}. The tidal forces experienced by the planetary systems due to stellar encounters as perturbations. In this frame of reference, the acceleration $\mathbf{a}(P_j)$ experienced by a planet $P_j$ is
\begin{eqnarray}
	\mathbf{a}(P_j) & = &\mathbf{a}_{\rm int}(P_j) + \mathbf{a}_{\rm tidal}\nonumber \\
	                & = &\mathbf{a}_{\rm int}(P_j) + \Big[\mathbf{a}_{\rm C}(P_j) - \mathbf{a}_{\rm C}(S)\Big] \quad .
	\label{eq:acc}
\end{eqnarray}
$\mathbf{a}_{\rm int}(P_j)$ is the acceleration experienced by planet $P_j$ due to the presence of its host star and the $n_p$ other planets in the system:
\begin{equation}
	\mathbf{a}_{\rm int}(P_j) = -G \left( \frac{M_s\mathbf{r}_j}{r_j^3} + \sum_{i=1}^{n_p;i \neq j} \frac{m_i\mathbf{r}_{ij}}{r_{ij}^3} \right) \quad ,
	\label{eq:acc_int_p}
\end{equation}
where $M_s$ is the mass of the host star, $m_i$ are the masses of the other planets, $\mathbf{r}_{j}$ is the position vector of the $j$-th planet, $\mathbf{r}_{ij} \equiv \mathbf{r}_{i} - \mathbf{r}_{j}$, and $G$ is the gravitational constant.

In a star cluster with $N$ stars, suppose that $\mathbf{R}_i$ is the position vector of the $i$-th star with respect to the cluster center. The acceleration experienced by the host star due to the other $N-1$ stars in the star cluster is then
\begin{equation}
	\mathbf{a}_{\rm C}(S) = -G \sum_{p=1}^{N; p \neq h} \frac{M_p\mathbf{R}_{ph}}{R_{ph}^3} \quad .
	\label{eq:acc_sun}
\end{equation}
where $\mathbf{R}_{ph} = \mathbf{R}_{p} - \mathbf{R}_{h}$ is the relative position vector from the host star $h$ to the perturber $p$, and $M_p$ is the mass of the perturber. Finally, the acceleration experienced by planet $P_j$ due to the other $N-1$ stars in the star cluster is calculated as
\begin{equation}
	\mathbf{a}_{\rm C}(P_j) = -G \sum_{p=1}^{N; p \neq h} \frac{M_p (\mathbf{R}_{ph} - \mathbf{r}_j)}{(|\mathbf{R}_{ph} - \mathbf{r}_j|)^3} \quad ,
	\label{eq:acc_sc_to_p}
\end{equation}

Consider a simplified scenario where the planetary system is perturbed by only one perturber of mass $M_p$. The tidal force in the vicinity of the host star is 
\begin{equation}
	\mathbf{\dot{a}}_{\rm C}(S) = \frac{d \mathbf{a}_{\rm C}(S)}{d \mathbf{R}_{ph}} = \frac{2 G M_{ph}}{|\mathbf{R}_{ph}|^3} \hat{\mathbf{R}}_{ph} \quad ,
\end{equation}
where ${\hat{\mathbf{R}}_{ph}}$ is the unit vector of $\mathbf{R}_{ph}$. Therefore, the tidal force is proportional to $R_{ph}^{-3}$ (for simplicity, hereafter we call it the $r^{-3}$ dependence), which is a strong function of the distance between the targeted planetary system and the perturber. If $r_j \ll R_{ph}$, to the first order of $\mathbf{r}_{j}$, the acceleration experienced by planet $P_j$ due to the perturber is
\begin{equation}
	\mathbf{a}_{\rm C}(P_j) = \mathbf{a}_{\rm C}(S) + \mathbf{\dot{a}}_{\rm C}(S) \mathbf{r}_{j} + \mathcal{O}(\mathbf{r}^2_{j}).
\end{equation}

In reality, planetary systems in a star cluster are perturbed simultaneously by $N-1$ member stars. Direct summation of the tidal forces due to these $N-1$ stars is possible but expensive. To explore whether it is possible to simplify the calculation by taking only the closest perturber into account, let us assume that $N$ stars are distributed in a virialized ($Q = 0.5$) \cite{plummer1911} sphere, whose gravitational potential exhibits the form:
\begin{equation}
	\Phi_{\rm P}(R) = - \frac{GM}{\sqrt{R^2+b^2}},
\end{equation}
where $b$ is the Plummer scale length, $M$ is the total mass of the Plummer sphere (i.e., the total mass of the cluster in our case), and $R$ is the distance to the cluster center. As such, around the cluster center where $R \sim 0$, the potential is roughly constant. The tidal force of the cluster as a function of $R$ is therefore
\begin{equation}
	|\ddot{\Phi}_{\rm P}(R)| = \frac{d^2 \Phi_{\rm P}(R)}{d R^2} = \frac{GM(b^2-2R^2)}{(b^2+R^2)^{5/2}},
\end{equation}
which reaches the maximum at the cluster center (i.e. $R = 0$). At this point, the relative magnitude of the cluster tide normalized to the magnitude of the perturber tide is 
\begin{equation}
	\frac{|\ddot{\Phi}_{\rm P}(R)|}{|\mathbf{\dot{a}}_{\rm C}(S)|} = \frac{GM}{b^3} \frac{R^3_{ph}}{2GM} = \frac{M}{m} \left( \frac{R_{ph}}{b} \right)^3 \approx N \left( \frac{R_{ph}}{b} \right)^3 .
\end{equation}
For a typical cluster, the virial radius $R_{\rm v} \sim 1$~pc. Since $b = (3\pi/16) R_{\rm v}$ \citep{heggie2003}, therefore $b \sim 0.6$~pc. For a close encounter, we adopt $R_{ph} \sim 1000$~AU \citep{adams06}, therefore $R_{ph} \ll b$. For this reason, we conclude that the perturbations experienced by a targeted planetary system in a star cluster is dominated by the contributions from its neighbor stars\footnote{In principle, the Galactic tide acting on the host star cluster is also acting on its planetary systems. In practice, however, we consider the Galactic tide in the solar neighborhood negligible for the evolution of planetary systems.}.

When solving the equation of motion of each planet (Eq.~\ref{eq:acc}), we use
\rebound{} \citep{rein12} to integrate the term $\mathbf{a}_{\rm int}(P_j)$ (Eq.~\ref{eq:acc_int_p}). The term $\mathbf{a}_{\rm C}(S)$ (Eq.~\ref{eq:acc_sun}) is obtained by summing the pairwise gravitational accelerations from the host star to other stars in the star cluster, which are integrated by \nbodylong{} \citep[][and the references therein]{wang2015a, wang2016,aarseth03, spurzem99}. The term $\mathbf{a}_{\rm C}(P_j)$ (Eq.~\ref{eq:acc_sc_to_p}) is rather cumbersome to compute, as it requires both the positions of each star in the cluster, and the position of each planet in the planetary systems. Ideally, one would obtain the positions of all stars from \nbodylong{} and sum up the pairwise accelerations from these stars to a given planet located at a specific position in its orbit. However, the dynamical timescales of star clusters ($\sim$ Myr) is many orders of magnitudes longer than that of planetary systems (down to a few days for hot Jupiters). Each planetary orbit requires $\sim$100 integration steps (depending on the actual integrator) to maintain reasonable accuracy. In each step, the combined tidal force due to other $N-1$ stars needs to be evaluated with respect to the current position of each planet. The coupled system of planetary systems in star clusters is therefore highly hierarchical. It is not practical to integrate such systems with the small timesteps required.

In an analogous scenario where a star cluster evolves in a galactic tidal field, tidal tensors are often used to describe the tidal effects \citep[e.g.,][]{renaud2011,rieder2013}, which turns out to be reliable and accurate. Despite this analogy, tidal tensors may not be directly applicable to planetary systems that are perturbed by stellar encounters, due to strong and rapid fluctuations in the tidal field. 

As mentioned above, neighbor stars dominate the perturbations of the planetary systems. This property allows us to dramatically reduce the computational costs substantially by considering only the perturbations due to the $n$ nearest neighbors, where $n \ll N$. In the case where $n=1$, only the nearest star is taken into account as the perturber, which is analogous to the Monte-Carlo scattering simulations, such as those of \cite{hao2013}, with the important difference that the perturbers are obtained from direct integration of the host star clusters where the multiple mechanisms such as dynamical evolution, stellar evolution, and galactic tidal fields can be taken into account. The case $n=0$ corresponds to an isolated planetary system, which we will use for validation and verification of our methods.

Throughout our study we adopt $n=1$. The mass of the nearest perturber and its position with respect to the planetary system under consideration are communicated to the planetary system integrator. The planetary system integrator subsequently calculates the tidal force experienced by all planets at each integration step. 

%%%%%%%%%%%%%%%%%%%%%%%%%%%%%%%%%%%%%%%%%%%%%%%%%%%%%%%
%%%%%%%%%%%%%%%%%%%%%%%%%%%%%%%%%%%%%%%%%%%%%%%%%%%%%%%
%%%%%%%%%%%%%%%%%%%%%%%%%%%%%%%%%%%%%%%%%%%%%%%%%%%%%%%

\subsection{Methods and Optimization}
Planetary systems and star clusters are two very different dynamical systems. Planetary systems are AU-scale few-body systems, whereas star clusters are parsec-scale many-body systems. The orbital periods of planets vary from several hours to a few hundred years, whereas the crossing timescale of star clusters are typically $\sim$~Myr. A simulation of planetary systems are generally considered satisfactory when the relative energy error is $|\Delta E|/E_0 \aplt 10^{-10}$, whereas in star cluster simulations this criterion is usually relaxed to $|\Delta E|/E_0 \aplt 10^{-5}$. If a star has only one planet, it is possible to integrate the planet as a low-mass binary companion using the regularization technique \citep[e.g.][]{ksreg65}. This technique, however, becomes inadequate in handling multiplanetary systems embedded in star clusters. 

It is therefore necessary to integrate these two types of systems using dedicated integrators that are optimized for for each of these dynamical problems, respectively. We employ \rebound{} \citep{rein12} for the integration of planetary systems using the IAS15 algorithm \citep{rein15}, which is optimized  for handling close encounters. We use the \nbodylong{} package (with a 4th-order Hermite algorithm) to integrate the star cluster. The perturbations to the targeted planetary system is computed according to the positions of the stars, obtained through the \nbodylong{} code. Since the tidal force of the perturber depends sensitively on its distance to the targeted planetary system (especially during a close encounter), it is crucial to evaluate the exact positions of the perturbers accurately, with time steps comparable to the planet integration time step ($\sim$ 1/100th of the shortest orbital period, or even shorter). In practice, it is difficult to determine the perturber's exact position with such small time steps, as the star cluster integration time steps are orders of magnitude longer than the planet orbit integration time step. Different stars are usually assigned different integration time steps according to their dynamical ``activities'' -- an integration scheme called \emph{individual time step scheme}, designed to reduce the $\mathcal{O}(N^2)$ computational complexity to $\sim \mathcal{O}(N^{4/3})$\,\footnote{The actual computational complexity depends on the density profile of the star cluster}. 

In order to bridge the apparent gap of time steps between planetary systems and star clusters, we store the star cluster simulation data, and subsequently make interpolation of the positions of all stars to accommodate the time steps required by the planetary system integration. \cite{farr12} propose the \emph{Particle Stream Data Format (PSDF)} scheme optimized for recording the output of general $N$-body simulations that exploit individual time steps. The PSDF scheme records the data incrementally only when a particle updates its scale, and thereby reduce the redundancy. The data is presented with the {\tt YAML}\footnote{\tt http://yaml.org/} format, which is highly human readable. Inspired by this idea, we have developed an adaptive storage scheme called {\em Block Time Step (BTS) storage scheme} to incrementally store star cluster data at arbitrarily high spatial and temporal resolutions \citep{cai2015}. The BTS scheme makes it possible to reconstruct the star cluster evolution with full details of stellar encounters with controllable snapshot sizes. To facilitate high-performance parallel access to the data files, we instead store the data with the \hdf{}\footnote{\tt https://www.hdfgroup.org/} data format. Accordingly, we carry out star cluster simulations separately and store the integration data with a temporal resolution comparable to 1000\,yr per snapshot. Dynamical data such as positions ($\mathbf{x})$, velocities ($\mathbf{v}$), accelerations ($\mathbf{a}$) and the first derivative of the accelerations ($\mathbf{\dot{a}}$) are stored for the purpose of interpolation. 

The coupling between star cluster dynamics and planetary system dynamics is implemented within the \amuse{} (Astrophysical Multipurpose Software Environment)\footnote{\texttt{http://amusecode.org/}} framework. We construct a GPU-accelerated pseudo-gravitational dynamics interface {\tt H5nb6xx}, which loads the BTS time series data stored in \hdf{} files. At $T=0$ when a simulation starts, {\tt H5nb6xx} reads the initial state of the star cluster, assigns an ensemble of initially identical multiplanetary systems to solar-type stars. Each planetary system is integrated by one \rebound{} instance implemented in the \amuse{} framework. The \rebound{} instances advance their own planetary systems, inquire the positions and masses of the nearby perturbers at time $T$. {\tt H5nb6xx} responds to these inquiries by loading two adjacent snapshots at $T_0$ and $T_1$, where $T_0 \leq T < T_1$. Accordingly, the particle states at $T$ are interpolated in parallel on the GPU using a set of seventh-order septic splines. Eventually, accurate positions of the host stars and their neighbors are obtained and transmitted into each \rebound{} instance, which carries out the integration of planetary systems with the additional forces from the perturbers. As such, {\tt H5nb6xx} and \rebound{} communicate iteratively until the simulation completes. During the simulation, the snapshot of the coupled system is stored at a fixed time intervals for the purpose of optional restarting. The simulations are carried out using \texttt{SiMon}\footnote{Available at: \href{https://github.com/maxwelltsai/SiMon}{https://github.com/maxwelltsai/SiMon}} (Qian et al. 2017, submitted), an open source Simulation Monitor for computational astrophysics.

A planet is identified as having escaped from its host star when its orbital eccentricity $e \geq 0.995$ during at least five consecutive integration time steps\footnote{Physically speaking, a planet is unbound only if its eccentricity $e \geq$ 1.0. In practice, however, this causes numerical difficulties for the integration, and therefore we relax this criterion to $e \geq 0.995$ in the last five consecutive time steps. }. In such a case, the planet is removed from the planetary system and becomes a free-floating planet (FFP). Depending on the escape velocity, the FFP may remain in the host star cluster and get recaptured \citep{wang2015b} or escape from the host star cluster \citep{zheng2015} .

%%%%%%%%%%%%%%%%%%%%%%%%%%%%%%%%%%%%%%%%%%%%%%%%%%%%%%%
%%%%%%%%%%%%%%%%%%%%%%%%%%%%%%%%%%%%%%%%%%%%%%%%%%%%%%%
%%%%%%%%%%%%%%%%%%%%%%%%%%%%%%%%%%%%%%%%%%%%%%%%%%%%%%%

\section{Initial Conditions}  \label{sec:ic}

\subsection{Initial Conditions for Star Clusters} \label{sec:sc_ic}
While the total number of stars in star clusters ranges between $10^2$ and $10^7$, the likelihood for dense globular clusters to bear planets is low \citep[see, e.g.,][]{Gilliland2000, gonzalez2001, madusa2017}. We therefore study intermediate-size open clusters consisting of $N=2\kilo, 8\kilo$ and $32\kilo$ stars, which are comparable with the total masses of \textsc{M67} \citep[e.g.,][]{hurley05}, \textsc{NGC 6811} \citep[e.g.,][]{meibom13} and the \textsc{Westerlund 1} \citep[e.g.,][]{portegies10} open clusters, respectively. The virial radii of all models are initialized at 1~pc, resulting in a range of central stellar densities that vary by a factor of four. We adopt a \cite{kroupa2001} stellar initial mass function with a mass range of $0.08-25~\msun$, which corresponds to a mean stellar mass of $0.52~\msun$. The stellar positions and velocities are sampled from the \citep{plummer1911} model in virial equilibrium (i.e., virial ratio $Q=0.5$). We do not include primordial binaries or primordial mass segregation. We evolve our models for 50~Myr, a time during which stellar evolution can be ignored for the mass range under consideration.

%%%%%%%%%%%%%%%%%%%%%%%%%%%%%%%%%%%%%%%%%%%%%%%%%%%%%%%
%%%%%%%%%%%%%%%%%%%%%%%%%%%%%%%%%%%%%%%%%%%%%%%%%%%%%%%
%%%%%%%%%%%%%%%%%%%%%%%%%%%%%%%%%%%%%%%%%%%%%%%%%%%%%%%

\subsection{Initial Conditions for Planetary Systems} \label{sec:p_sys_ic}

Current exoplanet data\footnote{see e.g., {\tt http://exoplanet.org}} show that planetary systems are immensely diverse: eccentricities are widely spread and the distributions of semi-major axis and mass seem to exhibit complex patterns. In this study we restrict ourselves to systems in which all planets have equal mass, are initially on coplanar and have circular orbits. We further assume that their semi-major axes are equally spaced in terms of their mutual Hill radii (dubbed \emph{EMS}: {\bf E}qual {\bf M}ass and equal {\bf S}eparation in terms of mutual Hill radii, see \citealt{zhou07,hao2013}). The scaled separation $k$ of the planetary orbits is
\begin{equation}
 k = \frac{a_{i+1} - a_{i}}{R_H} \quad ,
 \label{eq:k}
\end{equation} 
where $a_i$ is the semi-major axis of the $i$-th planet ($i = 1, 2,\dots,n$), and $R_H$ is the mutual Hill radius: 
\begin{equation}
 R_H = \left(\frac{2\mu}{3}\right)\frac{a_i + a_{i+1}}{2} \quad .
 \label{eq:r_h}
\end{equation} 
The quantity $\mu = m/M$ is the mass ratio between a planet and its host star.

\begin{table*}
	\begin{tabular}{lccclllll}
		\hline
		\hline
		Model & $k$ & $m_{\rm p}$ ($\mjup$) & $N_{\rm p}$ & $a_0$ [AU] & 2k & 8k & 32k & Remarks \\
		\hline
		\textsc{Model I}  & 10  & 1 & 5 & [1, 2.5, 6.3, 15.9, 39.7] & 50 & 50 & 200 & compact multiple-Jupiters \\
		\hline
		\textsc{Model II} & 100 & $10^{-3}$ & 5 & [1, 2.5, 6.3, 15.9, 39.7] & 50 & 50 & 200 & compact multiple-Earths\\
		\hline
		\textsc{Model III} & 10 & 1 & 5 & [5.2, 13.04, 32.7, 82.2, 206.2] & 50 & 50 & 200 & wide multiple-Jupiters\\
		\hline
		\textsc{Model IV} & 100 & $10^{-3}$ & 5 & [5.2, 13.04, 32.7, 82.2, 206.2] & 50 & 50 & 200 & wide multiple-Earths\\
		\hline
		\hline
	\end{tabular}
	\caption{Initial conditions of the planetary system models. The host stars are chosen from solar-type stars in the cluster. \textsc{Model I} and \textsc{Model II} are ``compact'' models, with the inner edge comparable to the Earth's orbit, and the outer edge comparable to Pluto's orbit. \textsc{Model III} and \textsc{Model IV} are the ``wide'' models, with the inner edge comparable to Jupiter's orbit, and the outer edge of $\sim 200$~AU comparable to $\sim 40\%$ of Sedna's orbital semi-major axis. \textsc{Model I} and \textsc{Model III} are multiple-Jupiter models where planet-planet interactions are important; \textsc{Model II} and \textsc{Model IV} are multiple-Earth model where where the gravitational interactions among planets can be safely ignored most of the time. The number of planetary systems in each ensemble are listed in the 2k, 8k, 32k columns, respectively. In total, there are 1200 individual planetary system simulations, and each simulation is carried out for 50~Myr.}
	\label{tab:planetary_ic}
\end{table*}

We study four EMS models as detailed in Table~\ref{tab:planetary_ic}. All models adopt $n=5$ planets orbiting around solar-type stars ($M = 1.00 \pm 0.02~\msun$). \cite{kokubo98} suggest that separations of $k \sim 10$ are typical, and therefore $k = 10$ is adopted in the multiple-Jupiter models (\textsc{Model I} and \textsc{Model III}). The inner edge of \textsc{Model I} is comparable to the Earth's orbit, while that of \textsc{Model III} is comparable to Jupiter's orbit. \textsc{Model II} and \textsc{Model IV} are obtained by reducing the planet mass of \textsc{Model I} and \textsc{Model III} by a factor of 1000 while keeping all other parameters unchanged (i.e., $m=10^{-3}~\mjup$ and $\mu \sim 10^{-6}$, thus comparable to $\sim 1/3$ Earth mass). According to Eq.~\ref{eq:k} and~\ref{eq:r_h}, the corresponding separation is $k = 100$ in this configuration. The $k = 100$ configurations serve as comparisons that can be used to disentangle the effects of stellar encounters and planet-planet interactions on the dynamical evolution of planetary systems in star clusters. The wide orbits in \textsc{Model III} and \textsc{Model IV} are useful to provide insights into how external perturbations shape the orbits of objects with large semi-major axes, such as trans-Neptunian objects (TNOs).

Since planetary systems are assigned to solar-type stars, their spatial distribution is random across the cluster. Given that the density profile of Plummer model follows $\rho(r) \sim r^{-5}$ \citep{plummer1911}, planetary systems in the central region of the star cluster are immersed in much higher stellar densities than their siblings in the outskirts, therefore they experience much stronger perturbations and more frequent encounters. Our simulations focus on the post-formation regime where the protoplanetary disks have already dissipated. Gas drag is therefore not taken into account. Tidal circularization is important when a planetary orbit is excited to high eccentricity and the periapsis with respect to the host star is small \citep{chatterjee08}. Tidal circularization is particularly important when the planetary system is subject to Lidov-Kozai cycles. This is a common mechanism for producing hot Jupiters \citep[e.g.,][]{shara2016, hamers17}. Tidal circularization damps the eccentricity of a perturbed planet, which protects the planet and its planetary system. However, in our $N$-body simulations we do not take this effect into account.

%%%%%%%%%%%%%%%%%%%%%%%%%%%%%%%%%%%%%%%%%%%%%%%%%%%%%%%
%%%%%%%%%%%%%%%%%%%%%%%%%%%%%%%%%%%%%%%%%%%%%%%%%%%%%%%
%%%%%%%%%%%%%%%%%%%%%%%%%%%%%%%%%%%%%%%%%%%%%%%%%%%%%%%

\section{Results} \label{sec:results}

\subsection{Statistics of Stellar Encounters}

Over the timespan of 50 Myr in our simulations, most cluster stars have experienced dozens of crossing times and roughly a half of relaxation time. Therefore, planetary systems within the star cluster will have experienced substantial number of encounters, depending on the neighbor density in the vicinity\footnote{In this paper, we define an encounter as the time span since a perturber becomes the closest star to the targeted planetary system, until it is replaced by another star even closer to the planetary system. In other words, each change of neighbor star is considered as an encounter, regardless of its strength or duration.}. Inspired by the approach of modeling the encounter between the host star and a external perturber with a two-body problem \citep[e.g.,][]{spurzem09,heggie2006}, we carry out our analysis with such a model, and quantify the strength of \emph{each} encounter with dimensionless parameters $\vinf$ and $K$. The quantity $\vinf$ is the relative speed of the perturber with respect to the host star, scaled to the average orbital speed of the outermost planet. The parameter $K$ is the ratio of the perturbation timescale to the planet orbital timescale, defined as
\begin{equation}
	K = \sqrt[]{\frac{2 M_{\rm s}}{M_{\rm s} + M_{\rm p}} \left( \frac{r_p}{a} \right)^3} \quad ,
	\label{eq:K_encounter}
\end{equation}
where $M_{\rm s} = m_{\rm star} + m_{\rm pl}$ is the total mass of the perturbed planetary system, $M_{\rm p}$ is the mass of the perturber, $r_p$ is the pericenter distance of the perturber with respect to the planetary system center of mass, and $a$ is the semi-major axis of the perturbed planetary orbit.

%Figure~\ref{fig:K_vinf} shows the ensembles of encounters for $N = 1\kilo$ and $N = 8\kilo$ hosting clusters in the $(K - \vinf)$ space. In each panel, two species of encounters are identified: hyperbolic or parabolic flybys when $\vinf \geq 0$, and $\vinf<0$ otherwise. The frequency distribution of $\log K$ is shown in Figure~\ref{fig:K_freq_normed}, and can be approximated with a log-normal distribution.

\begin{figure*}
\centering
\includegraphics[scale=0.8]{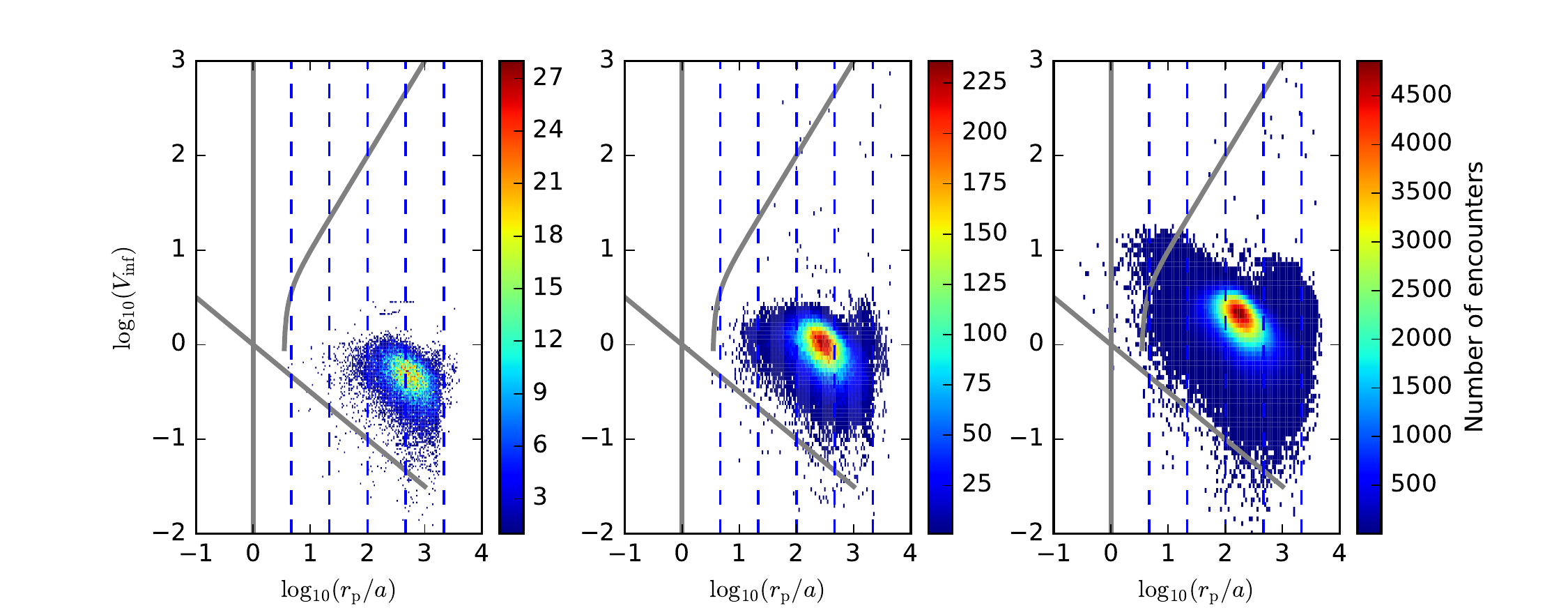}
\caption{Ensembles of stellar encounter parameters at the periapsis for the $N = 2\kilo, 8\kilo$ and $32\kilo$ star clusters (from left to right). The vertical gray line coincides with $r_{\rm p} = a$, where $a$ is the semi-major axes of the outermost planets ($\sim 40$~AU). If $r_{\rm p} \gg a$, the encounter is \emph{tidal}. The diagonal grey line coincides with the hyperbolic eccentricity $e'=1$. Above this line encounters are \emph{hyperbolic}, below the line they are \emph{nearly parabolic}. The grey curve separates the regime of \emph{adiabatic encounters} ($V_{\rm p} \ll v_{\rm cir, a}$, where $v_{\rm cir, a}$ is the circular velocity at $a$) and \emph{impulsive encounters} ($V_{\rm p} \gg v_{\rm cir, a}$). The five blue vertical dashed lines from left to right correspond to $K=10,10^2,10^3,10^4$ and $K=10^5$, respectively (assuming that the perturber mass is $M_p = 1 M_{\oplus}$, and the mass of the host star is $M_{\rm star} = 1.0M_{\oplus}$ according to the initial conditions). The colors of dots in each panel correspond to the counts of encounters. The encounter parameters are collected when the perturber reaches $r_{\rm p}$.}
\label{fig:K_vinf}
\end{figure*}

Figure~\ref{fig:K_vinf} shows the distances (scaled to the semi-major axes of the outermost planet) and the velocities (scaled to the velocity at infinity of the perturber) of three ensembles of stellar encounters in the $N=2\kilo,8\kilo$ and $32\kilo$ clusters during the 50\,Myr timespan of simulations whenever a perturber reaches the periapsis.

\begin{figure}
\centering
\includegraphics[scale=0.4]{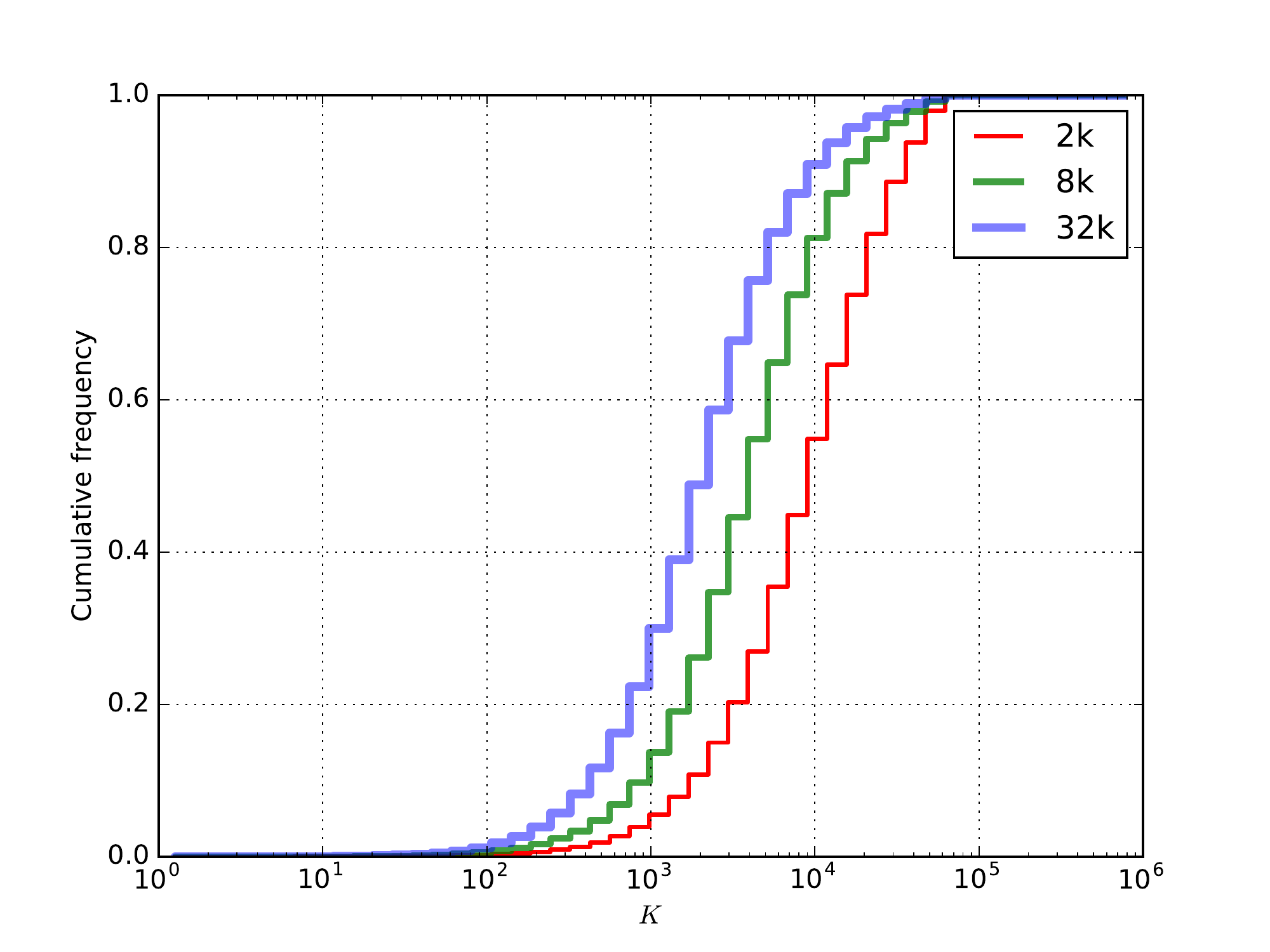}
\caption{Cumulative frequency distribution of the encounter strength parameter $K$ for the $N = 2\kilo, 8\kilo$ and $32\kilo$ hosting clusters. The cumulative curves shift leftward as the total number of particles $N$ grows higher, which shows that strong encounters are more common in denser star clusters. }
\label{fig:K_freq_normed}
\end{figure}

The frequencies of encounters (including nearly parabolic, hyperbolic, impulsive, tidal encounters; separated by the grey curves in the figure) increase as $N$ increases, indicating that more frequent encounters are expected in denser cluster environments. Indeed, as all our modeled clusters have identical $R_v = 1\textrm{ pc}$ initially, a larger $N$ results in higher stellar density, and consequently smaller $K$ since $K\propto (r_p/a)^{3/2}$ (see Eq. \ref{eq:K_encounter}). The strengths of encounters are indicated with the five dashed vertical lines in each panel, in which smaller $K$ values are associated with stronger encounters. As compared with hyperbolic encounters, the near parabolic regime is more destructive to the targeted planetary systems -- binary systems are formed between the perturber and the host star, with the possibility of triggering Kozai-Lidov oscillations \citep{naoz2011,hamers16}. The cumulative frequency distribution of $\log K$ is shown in Figure~\ref{fig:K_freq_normed}. The cumulative frequency spectra shift leftward with the increment of $N$, showing that the average strengths of encounters increases in dense stellar environments.

We will see below, that the hyperbolic regime, albeit weak, affects the planetary systems cumulatively by series of subsequent relatively weak to moderate encounters ($K<50$). Stars in their clusters experience Coulomb-like scattering -- the contributions of rare and strong encounters are comparable to the contributions of the cumulative effect of a series of frequent and weak encounters.

%%%%%%%%%%%%%%%%%%%%%%%%%%%%%%%%%%%%%%%%%%%%%%%%%%%%%%%
%%%%%%%%%%%%%%%%%%%%%%%%%%%%%%%%%%%%%%%%%%%%%%%%%%%%%%%
%%%%%%%%%%%%%%%%%%%%%%%%%%%%%%%%%%%%%%%%%%%%%%%%%%%%%%%

\subsection{Perturbed Planetary Systems}
Multiplanetary systems are fragile when experiencing stellar encounters in the star cluster environments \citep[e.g.,][]{portegies15, hao2013}. The planet survival rates depend on the frequency of close stellar encounters, which in turn depends on the stellar density. Planetary systems close to the dense cluster center are more frequently perturbed than those in the cluster outskirts. 

\begin{figure*} 
\centering
\includegraphics[scale=0.8]{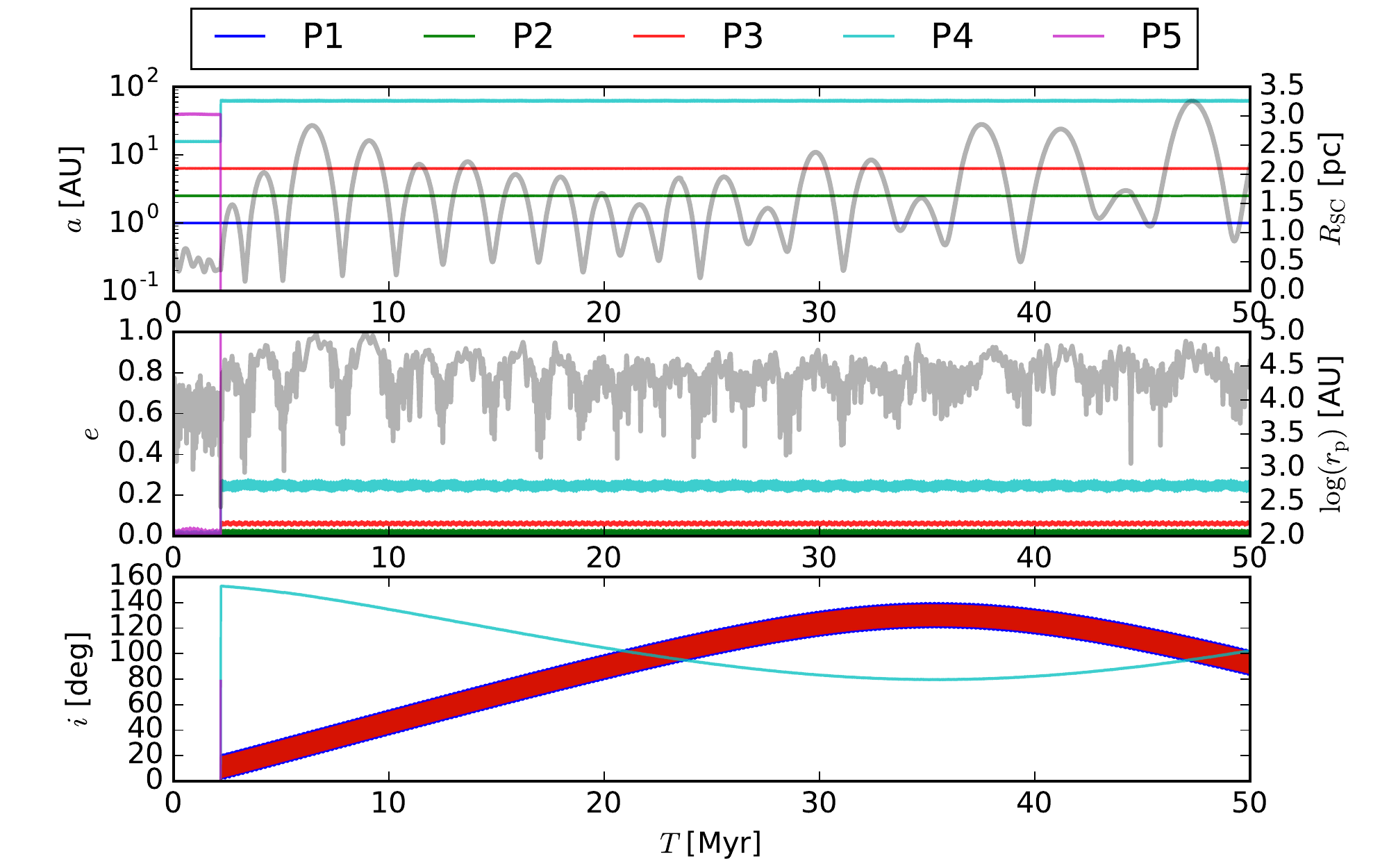}
\caption{Orbital element evolution of a \textsc{Model I} planetary system in the $N=8\kilo$ cluster. The system has 5 planets initially, which are initially on circular and coplanar orbits. In the top panel, the semi-major axis of each planet is plotted in colored curved as a function of time (in log-scale), while the thick gray curve indicates the distance from the host star to the cluster center (axis on the right, in parsec); in the middle panel, the evolution of the eccentricity of each planet is plotted as a function of time, and the thick gray curve is the distance of the perturber to the host star (in AU, log-scale); at the bottom panel the inclination of each planet with respect to the initial orbital plane is plotted, and the thick gray curve is the acceleration due to the perturber (log-scale, normalized to the acceleration due to the host star). In this particular planetary system, a close encounter encounter occurs at about $T=2.2$~Myr, which ejects the outermost planet (P5) immediately, and excites P4 to $e \sim 0.2$. This close encounter results in a prograde orbit ($i>90^{\circ}$) of P4, and also strengthens the planet-planet scattering among the remaining planets.}
\label{fig:ecc_rsc}
\end{figure*}

\begin{figure*}
\centering
\includegraphics[scale=0.8]{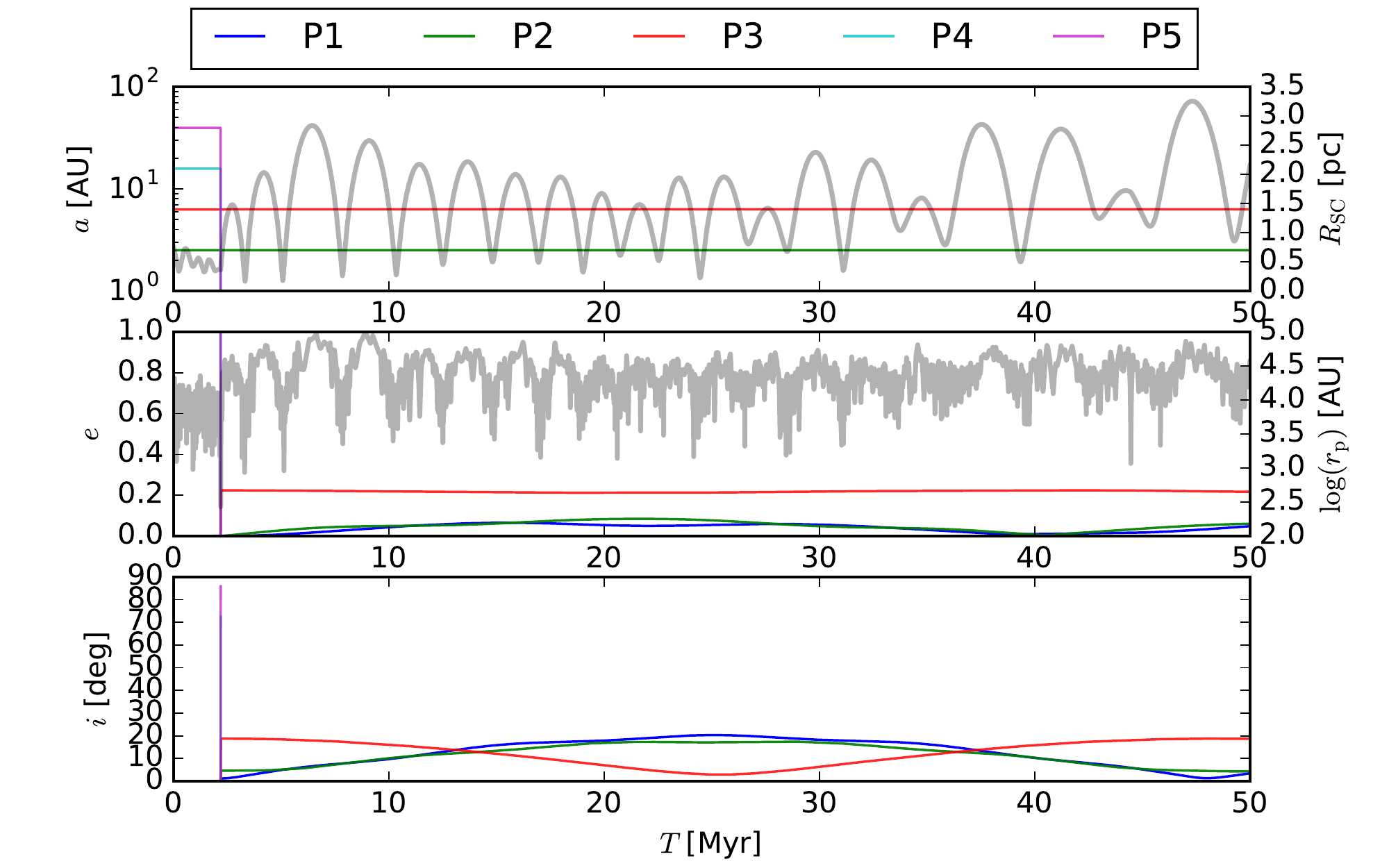}
\caption{The same host star in the same cluster as in Figure~\ref{fig:ecc_rsc}, but for a \textsc{Model II} planetary system. P4 and P5 are ejected immediately.}
\label{fig:ecc_rsc_2}
\end{figure*}

\begin{figure*}
\centering
\includegraphics[scale=0.8]{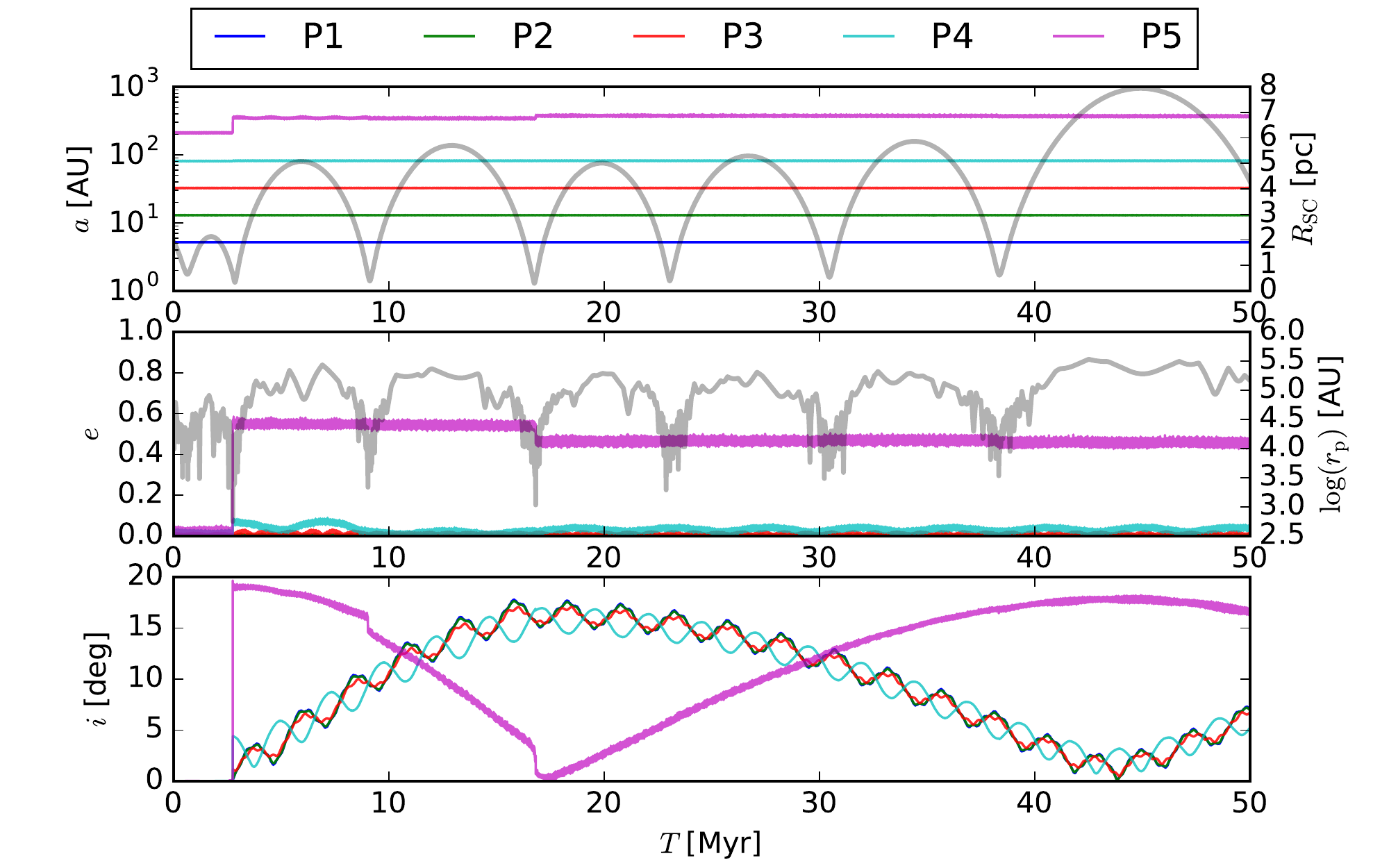}
\caption{Same as Fig.~\ref{fig:ecc_rsc_2}, but for a \textsc{Model III} planetary system in the $N=8\kilo$ cluster. The planetary system orbits around the cluster center in approximately eccentric orbits. As it dives into the cluster center, the frequency of perturbations increases significantly. Likewise, the planetary system remains roughly unperturbed when in the outskirts of the cluster. Consequently, excitation are more likely to occur during around the dense cluster center.}
\label{fig:ecc_rsc_3}
\end{figure*}

\begin{figure*}
\centering
\includegraphics[scale=0.8]{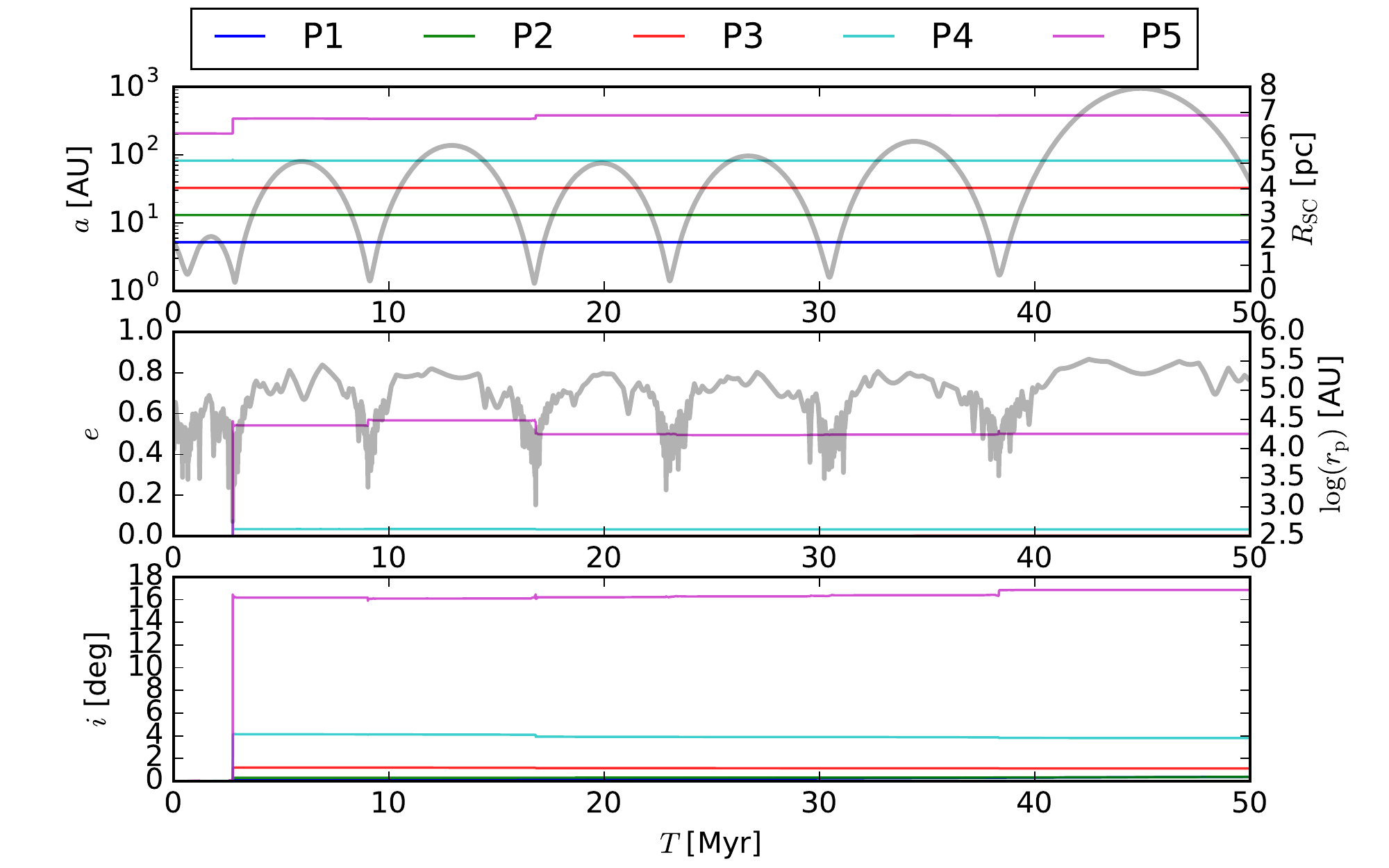}
\caption{Same as Fig.~\ref{fig:ecc_rsc_3}, but for a \textsc{Model IV} planetary system in an $N=8\kilo$ cluster.}
\label{fig:ecc_rsc_4}
\end{figure*}

Figures (\ref{fig:ecc_rsc} - \ref{fig:ecc_rsc_4}) depict the ``microscopic'' behaviors of four perturbed planetary systems. The first planetary system, shown in Figure~\ref{fig:ecc_rsc}, is a \textsc{Model I} multiple-Jupiter system in the $N=8\kilo$ host cluster, and its evolution is significantly affected by a perturbation at $T\sim 2.2$~Myr. The outermost planet P5 is immediately ejected, and the second outermost planet undergoes substantial eccentricity excitation and semi-major axis expansion. As such, the perturber exchanges energy and angular momentum with this planetary system, leading P4 into a retrograde orbit, and P1-P3 tightly coupled as a whole in the inner region of the system (as shown at the bottom panel in particular). For comparison, Figure~\ref{fig:ecc_rsc_2} presents the behavior of a \textsc{Model II} multiple-Earth planetary system orbiting exactly the same host star in the same cluster of Figure~\ref{fig:ecc_rsc}. The same close encounter at $T\sim 2.2$~Myr liberates P4 and P5 simultaneously. However, the system does not exhibit apparent pattern of planet-planet interaction as seen in Figure~\ref{fig:ecc_rsc}. Rather, planet are evolving mostly independently before and after the close encounter. In the similar way, the evolution of a \textsc{Model III} and a \textsc{Model IV} planetary systems are shown in Figure~\ref{fig:ecc_rsc_3} and Figure~\ref{fig:ecc_rsc_4}, respectively. The wide orbits of P5 in these two models are so sensitive to external perturbations that even the mild excitation of eccentricities can be used to trace the history of weak stellar encounters \citep[cf.][]{portegies15}. In both models, eccentricity excitations occur almost exclusively when the planetary system plunges into the dense cluster center, a region where the frequency and strength of encounters significantly increase. Both Figure~\ref{fig:ecc_rsc} and Figure~\ref{fig:ecc_rsc_3} exhibit interesting behavior of inclination evolution: slow anti-phase variation of inclinations are seen between the outermost planets and the inner planets combined; the inner planets undergo more rapid anti-phase oscillation of orbital inclinations among themselves. This behavior suggests that the secular evolution of planetary system is largely unaffected by weak encounters, but they can be dramatically changed by strong encounters through the injection of orbital energy and angular momentum.

These are only a few examples of the intricate evolution of planetary systems due to both internal and external perturbations. In general we observe in our simulations that these effects depend on the orbit of the planetary system in the star cluster. We can however, deduce several general behaviors. Shortly after the dissipation of protoplanetary disks, planets have obtained nearly circular and coplanar orbits. External perturbations initialized by stellar flybys pump up the eccentricities of planetary orbits. The growth in eccentricity $\delta e$ is proportional to the semi-major axes $a$ as $\delta e \sim a^2$ \citep[][Eq.~13]{spurzem09}, so outer planets with wide orbits are more vulnerable. Indeed, outer planets are excited more quickly, as can be seen in Figures (\ref{fig:ecc_rsc} - \ref{fig:ecc_rsc_4}). As the eccentricity of the outer planets grows due to external perturbations, its orbit intersects with the orbits of inner planets. Orbit crossings lead to planet-planet scattering, which in turn result in an inward propagation of eccentricity. This suggests that multiplicity does contribute to the vulnerability of planetary systems. As such, we predict that the stability of a planetary system can be compromised if there exist massive planets with large semi-major axes.  Interestingly, while the outermost planets are generally most vulnerable to external perturbations, they may not necessarily be the first ones to get ejected, because the excitation process is complicated by planet-planet interactions and orbital phases.

Moderate encounters ($10 < r_p/a < 30$), albeit weak, can still leave marks on the targeted planetary systems (see especially Figure~\ref{fig:ecc_rsc_3}). Each encounter results in a small $\delta e$ and $\delta a$, causing the orbit to gradually depart from its initial circular and coplanar state, a state corresponding to the maximum possible orbital angular momentum. The orbital angular momentum is partially taken away during the onset of an encounter, and therefore results in the growth of angular momentum deficit (AMD) \citep[see,][]{laskar97}. The AMD of outer planets is partially absorbed by inner planets, consequently causing the inner planets to develop eccentric orbits \citep{davies14}. As such, eccentricities are propagated from outer planets to inner planets. Due to the Coulomb-like random scattering among stars in the cluster, the frequency of moderate encounters (as seen in Figure~\ref{fig:K_vinf}) much higher than strong encounters ($r_p/a<10$), especially in denser clusters. These encounters are able to gradually pump up the AMD of a targeted planetary system by repeatedly perturbing the outermost planet. The value of AMD limits the maximum eccentricity and inclination an individual planet can attain. When the AMD is sufficiently large, the planetary system tends to reach an equipartition of AMD \citep{wu11}. Inner planets with small semi-major axis have relatively small orbital angular momentum, and therefore it only contribute to a small fraction of the total AMD. As a consequence of the AMD equipartition, the inner planet is ``forced'' to contribute as much AMD as possible, which in turn be driven to extreme orbits. It is worthy to point out that absorbing AMD takes time, and therefore if an inner planet is indirectly ejected because it is perturbed by an outer orbits, the ejection may happen well after the perturber departs from the periapsis. Additionally, an inner planet may be ejected earlier than the outer planets, because it has lower AMD capacity. Eventually, all planets in the system obtain high eccentricities, which results in the loss of stability of the entire system. 

We conclude that planets can be liberated immediately through very strong encounters ($k < 10$). An alternative channel to eject planets is through the cumulative effect of multiple moderate encounters ($10 < r_p/a < 30$). Distance encounters ($r_p/a$ > 50) have no direct implication to stability. Furthermore, planet-planet interactions are the catalysts of the destruction of a secularly interacting planetary system, such as in our multiple-Jupiter models.

\subsection{Statistical Behavior of Perturbed Planetary System Ensembles}
% 1) Survival rates as a function of time ==> compare 
% 2) Individual survival rates
% 3) a-e space and a-i space
% 4) de-fraction histogram
% 5) a-rp final distribution ==> hot Jupiters?

\begin{figure*}
\centering
\includegraphics[scale=0.8]{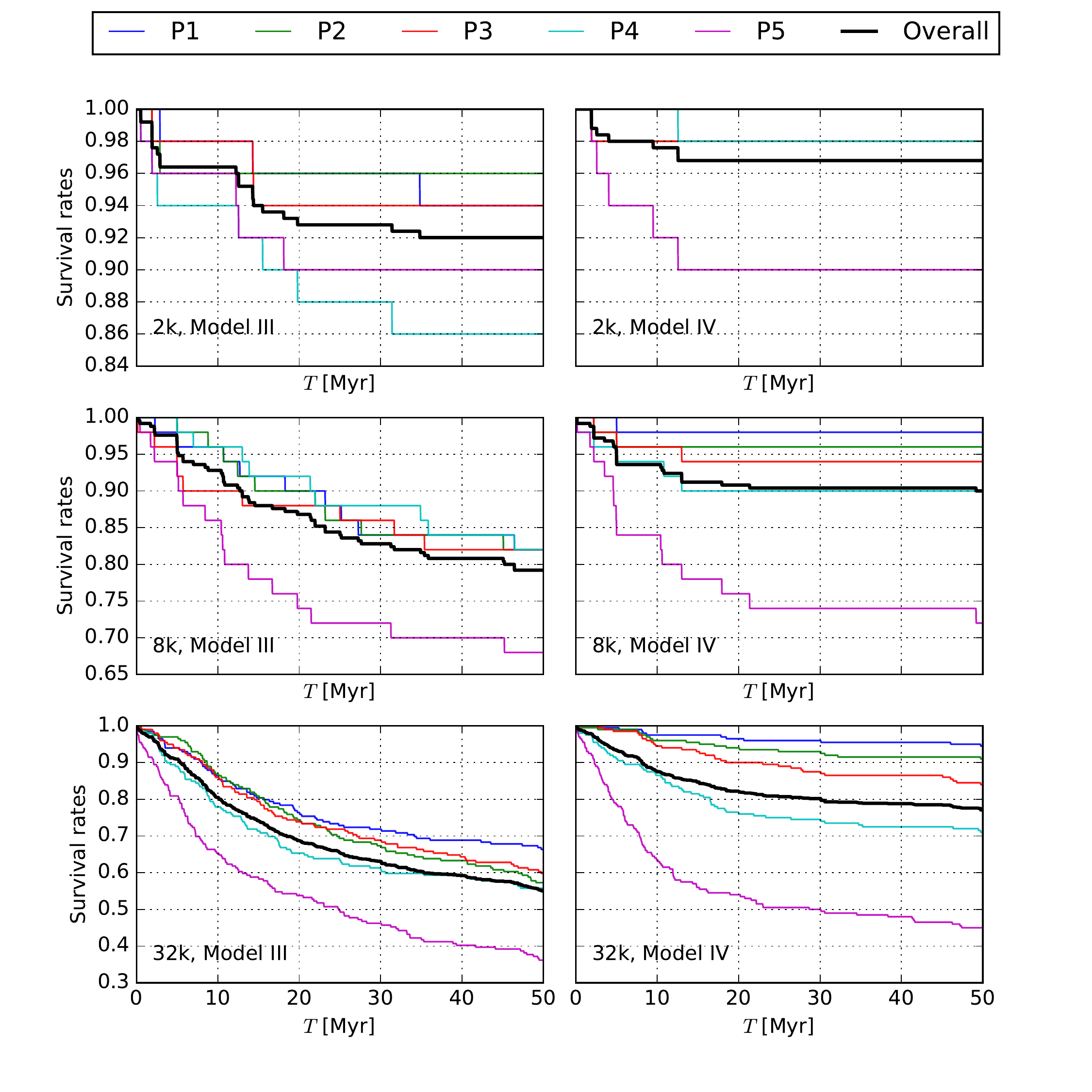}
\caption{Survival rates as a function of time for planetary systems in $N=2\kilo$ (top row), $N=8\kilo$ (middle row), and $N=32k\kilo$ clusters. The three panels on the left column are \textsc{Model III} planetary systems, and the three panels on the right column are \textsc{Model IV} planetary systems. Different planets are distinguished with different colors, and the thick black curve is the overall survival rates, defined as the ratio between the total number of ejected planets at the current time $T$ and the total number of of planets at $T=0$.  }
\label{fig:survival_rates}
\end{figure*}

Planetary systems are chaotic few-body systems sensitive to the initial conditions. In order to determine the contributions of external effects (stellar encounters) and internal effect (planet-planet interactions), we plot the planet survival rates\footnote{Survival rates $\eta$ of planets in a given host star cluster: defined as $\eta(T) = n_{\rm bp}(T)/n_{\rm bp}(T=0)$, where $n_{\rm bp}(T)$ is the total number of bound planets at time $T$.} as a function of time for different ensembles of simulations. As shown in Figure~\ref{fig:survival_rates}, six ensembles of simulations are plotted in their corresponding panels, with each of the planets distinguished with different colors. The final survival rates of each ensemble of simulations are listed in Table~\ref{tab:overall_survial_rates}.  Apparently, for both multiple-Jupiter models (\textsc{Model I} and \textsc{Model III}) and multiple-Earth models (\textsc{Model II} and \textsc{Model IV}), planetary systems in denser stellar environments suffer from higher ejection rates. When keeping the initial arrangement of semi-major axis fixed, the survival rates of multiple-Earth models are substantially higher. When keeping the initial orbital separation $k$ fixed, the compact orbit models (\textsc{Model I} and \textsc{Model II}) have significantly higher survival rates compared to the wide orbit models (\textsc{Model II} and \textsc{Model IV}).

Therefore, we can conclude that both internal effect and external effect play important roles in the evolution of planetary systems in star clusters. Furthermore, outer planets tend to be ejected more rapidly; tight orbit inner planets are better protected against ejections.

\begin{table}
	\centering
	\begin{tabular}{lllll}
		\hline
		\hline
		Models & \textsc{Model I} & \textsc{Model II} & \textsc{Model III} & \textsc{Model IV} \\
		\hline
		2k  & 0.984 & 0.988 & 0.920 & 0.968 \\
		\hline
		8k  & 0.940 & 0.968 & 0.792 & 0.900 \\
		\hline
		32k & 0.834 & 0.916 & 0.550 & 0.771 \\
		\hline
		\hline
	\end{tabular}
	\caption{The overall survival for each ensemble of simulations at $T=50$~Myr.}
	\label{tab:overall_survial_rates}
\end{table}

\begin{figure}
\centering
\includegraphics[scale=0.58]{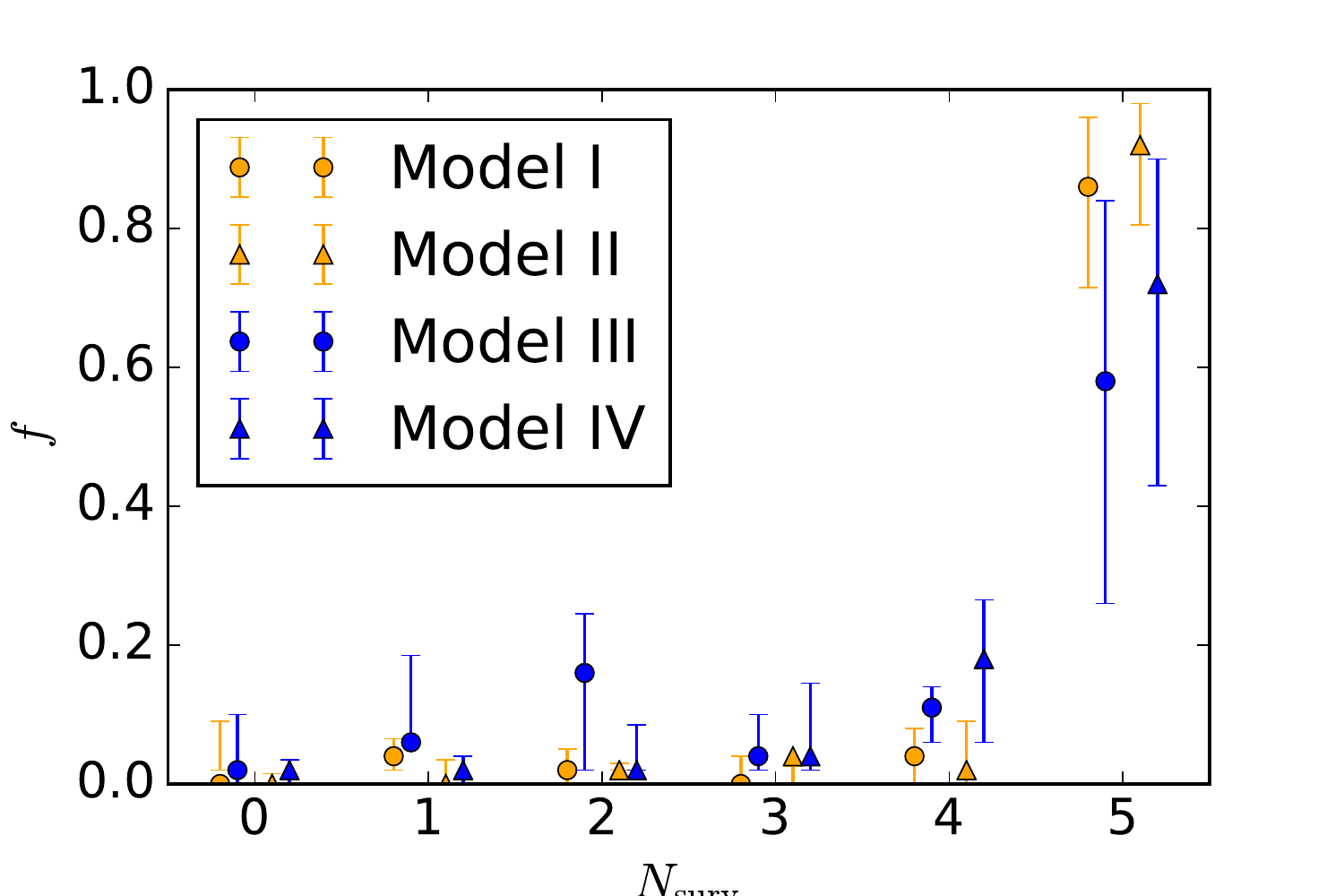}
\caption{Fractions of planetary systems $f$ as a function of the number of surviving planets in each system $N_{\rm surv}$ at $T = 50$~Myr. The data is from Table~\ref{tab:n_p_survived}. Circles: multiple-Jupiter models; triangles: multiple-Earth models; green: compact models; blue: wide models.}
\label{fig:mj_me_4k_8k_n_p_survived}
\end{figure}

Table~\ref{tab:n_p_survived} presents the fractions of planetary systems with each of them having exactly $n_{\rm surv}$ surviving planets ($0 \leq n_{\rm surv} \leq 5$) by the end of the simulation ($T=50$ Myr). A comparison between these fractions for the multiple-Jupiter and multiple-Earth models in the $N=8\kilo$ and $N=32\kilo$ clusters is presented in Figure~\ref{fig:mj_me_4k_8k_n_p_survived}. 
\begin{table}
	\centering
	\begin{tabular}{c|c|c|c|c|c|c|c}
		\hline
		\hline
		\multirow{2}{*}{$N$} & \multirow{2}{*}{Model} & \multicolumn{6}{c}{$N_{\rm surv}$ up to $T=50$~Myr} \\
		      & & {\bf 0} & {\bf 1} & {\bf 2} & {\bf 3} & {\bf 4} & {\bf 5} \\
		\hline
		$2\kilo$ & \textsc{I} &  0.02 &  0.02 & 0.0 & 0.0 & 0.0 & 0.96 \\
		\hline
		$2\kilo$ & \textsc{II} &  0.00 &  0.00 & 0.02 & 0.0 & 0.0 & 0.98 \\
		\hline
		$2\kilo$ & \textsc{III} &  0.00 &  0.06 & 0.02 & 0.02 & 0.06 & 0.84 \\
		\hline
		$2\kilo$ & \textsc{IV} &  0.00 &  0.00 & 0.02 & 0.02 & 0.06 & 0.90 \\
		\hline
		$8\kilo$ & \textsc{I} &  0.0 & 0.04 & 0.02 & 0.0 & 0.08 & 0.86 \\
		\hline
		$8\kilo$ & \textsc{II} &  0.0 & 0.0 & 0.02 & 0.04 & 0.02 & 0.92 \\
		\hline
		$8\kilo$ & \textsc{III} &  0.02 & 0.06 & 0.16 & 0.04 & 0.14 & 0.58 \\
		\hline
		$8\kilo$ & \textsc{IV} &  0.02 & 0.02 & 0.02 & 0.04 & 0.18 & 0.72 \\
		\hline
		$32\kilo$ & \textsc{I}  &  0.09 & 0.065 & 0.05 & 0.04 & 0.04 & 0.715 \\
		\hline
		$32\kilo$ & \textsc{II}  &  0.015 & 0.035 & 0.03 & 0.025 & 0.09 & 0.805 \\
		\hline
		$32\kilo$ & \textsc{III}  &  0.1 & 0.185 & 0.245 & 0.1 & 0.11 & 0.265 \\
		\hline
		$32\kilo$ & \textsc{IV}  &  0.035 & 0.04 & 0.085 & 0.145 & 0.265 & 0.43 \\
		\hline
		\hline
	\end{tabular}
	\caption{Fraction of planetary systems with the given number of planets survived in each planetary system $N_{\rm surv}$, at the end of each simulation ($T = 50$ Myr). }
	\label{tab:n_p_survived}
\end{table}

\begin{figure*}
\includegraphics[scale=0.9]{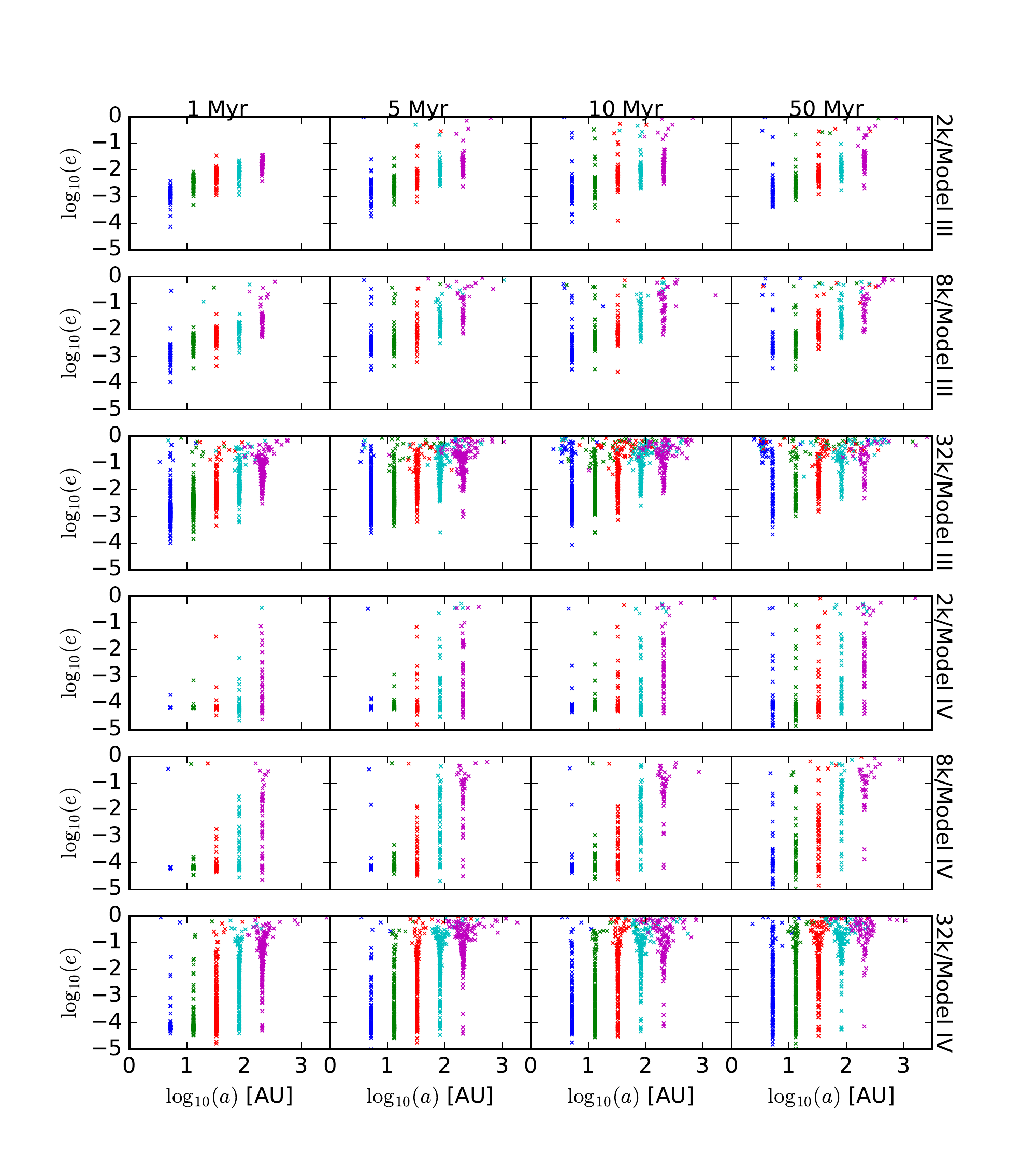}
\caption{Time frames of the $a - e$ space at $T=$~1~Myr (col.1), 5~Myr (col.2), 10~Myr (col.3) and 50~Myr (col.4). The 1-3 rows are for the \textsc{Model III} wide multiple-Jupiter planetary systems with hosting cluster of $N=2\kilo, 8\kilo$, and $32\kilo$, respectively. The 4-6 rows are for the \textsc{Model IV} wide multiple-Earth models in $N=2\kilo, 8\kilo$, and $32\kilo$ clusters, respectively.  }
\label{fig:ae_snapshots}
\end{figure*}

\begin{figure*}
\includegraphics[scale=0.9]{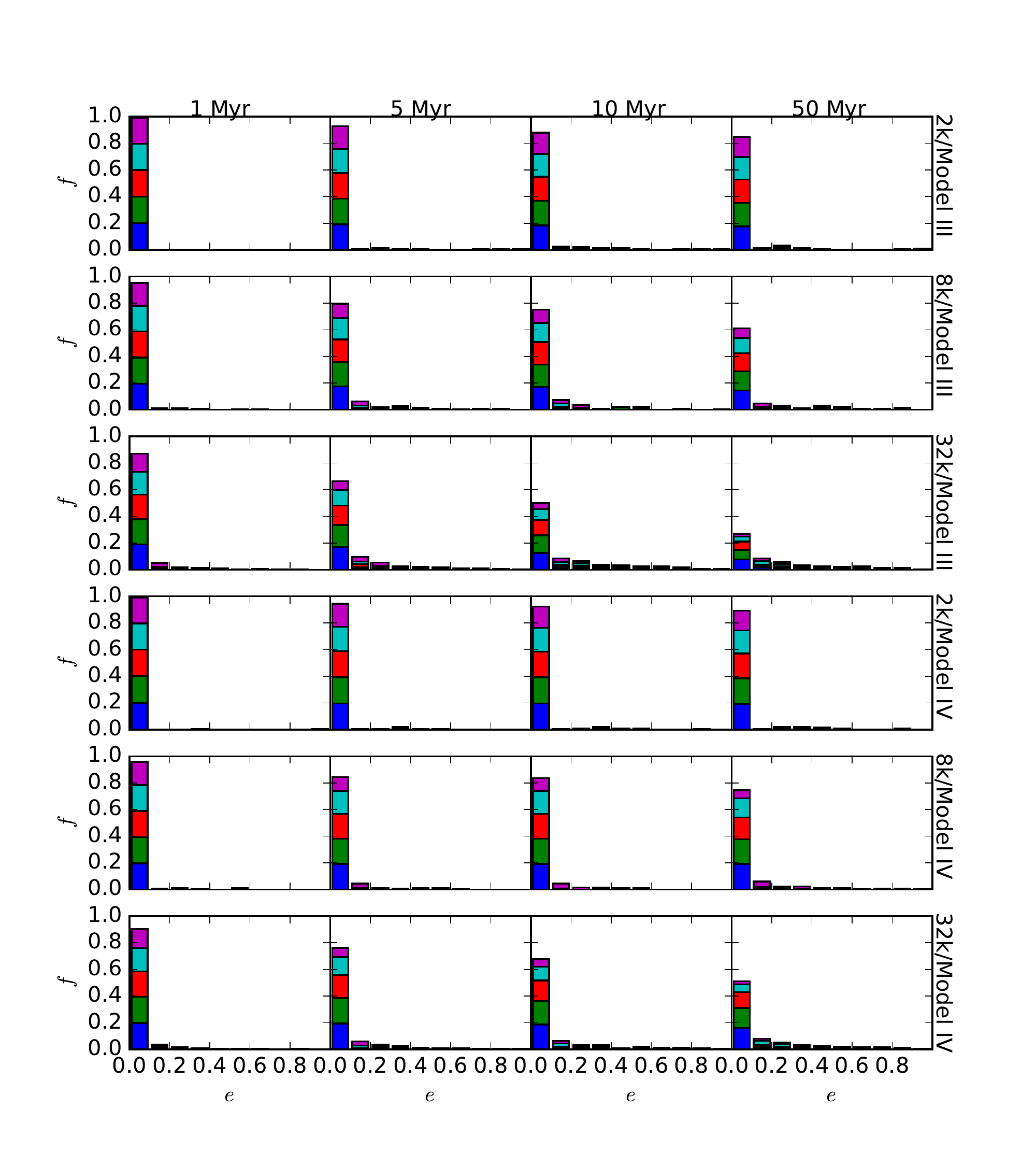}
\caption{Same with Figure~\ref{fig:ae_snapshots}, but for the distribution of eccentricities. }
\label{fig:ecc_dist_snapshots}
\end{figure*}

The excitation process of the ensembles of planetary systems can be inspected with Figure~\ref{fig:ae_snapshots}, where a grid of snapshots at the $a-e$ space is presented at four time checkpoints ($T=$~1, 5, 10 and 50~Myr) for different models. Planet migrations (changes in $a$) are seen especially among the planets in the multiple-Jupiter models with moderate to high eccentricities. For comparison, the corresponding $a-e$ snapshots of the multiple-Earth models are shown in the three bottom panels, where planet migrations are less pronounced. The population of highly-eccentric planets increases as $a$ increases, until the targeted planetary systems have experienced sufficient perturbations to even induce the eccentricity of the innermost planet, or until the AMD has reached an equipartition across the entire planetary system. The distribution of eccentricities at $T=50$~Myr is presented in Figure~\ref{fig:ecc_dist_snapshots}. 

\begin{figure}
\centering
\includegraphics[scale=0.42]{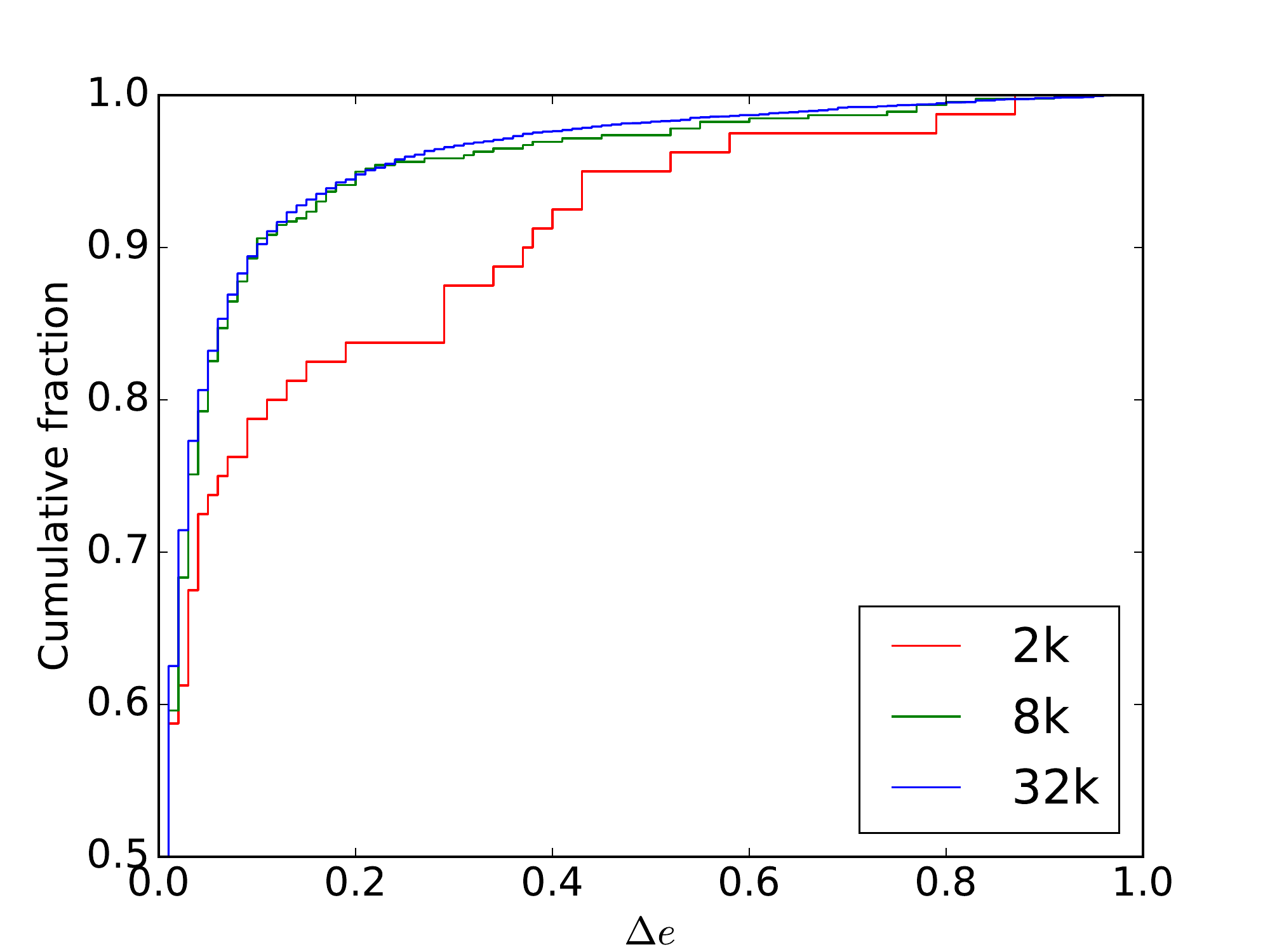}
\caption{The cumulative frequency of encounters that lead to a given change of eccentricity $\leq \Delta e$. Only $\Delta e \geq 0.01$ are plotted. The three curves correspond to three \textsc{Model IV} ensembles in $N=2\kilo,8\kilo$ and $32\kilo$ clusters, respectively. Weak encounters with small $\Delta e$ dominant the frequency spectra for all models. Very strong encounter causing $\Delta e \geq 0.5$ are rare. Due to the small number of close encounter events, the $N=2\kilo$ result is different from the the $N=8\kilo$ and $N=32\kilo$ results. }
\label{fig:de_fraction}
\end{figure}

We follow the changes in the orbital eccentricities $\Delta e \equiv |e_{\rm new} - e_{\rm old}|$ as a consequence of each stellar encounters. As shown in Figure~\ref{fig:de_fraction}, very strong encounters are rare. Among all models, only $\sim 3\%$ of encounters are sufficiently strong to cause excitations of $\Delta e \geq 0.5$. This suggests that planet ejections is a cumulative process -- planets are gradually excited by a number of subsequent encounters with relative small $\Delta e$, except for a few very strong encounters that ionize the planets immediately. These findings are consistent with the results in \citealt[][]{spurzem09} and \citealt[][]{hao2013}.

Hot Jupiters (HJs) can be produced with the combined effects of stellar encounters and planet-planet interactions. In the $N=32\kilo$ cluster, $\sim 0.2\%$ of \textsc{Model I} planets have developed orbital features of $r_{\rm p} \leq 0.1$~AU and $a \leq 1$~AU. Tidal circularization can be efficient when these planets are around their orbital periapsis, which in turn provides a mechanism to produce HJs.  The HJs rate predicted in \citep{shara2016} is $\sim 1\%$, higher than our results. However, we note that our simulations (50~Myr) are much shorter than the simulations in \citep{shara2016} (1~Gyr). We therefore suspect that if we were not restricted by the computational costs, our results will be more consistent with \citep{shara2016} if we carry out our simulations for longer time.

\begin{figure*}
\includegraphics[scale=0.9]{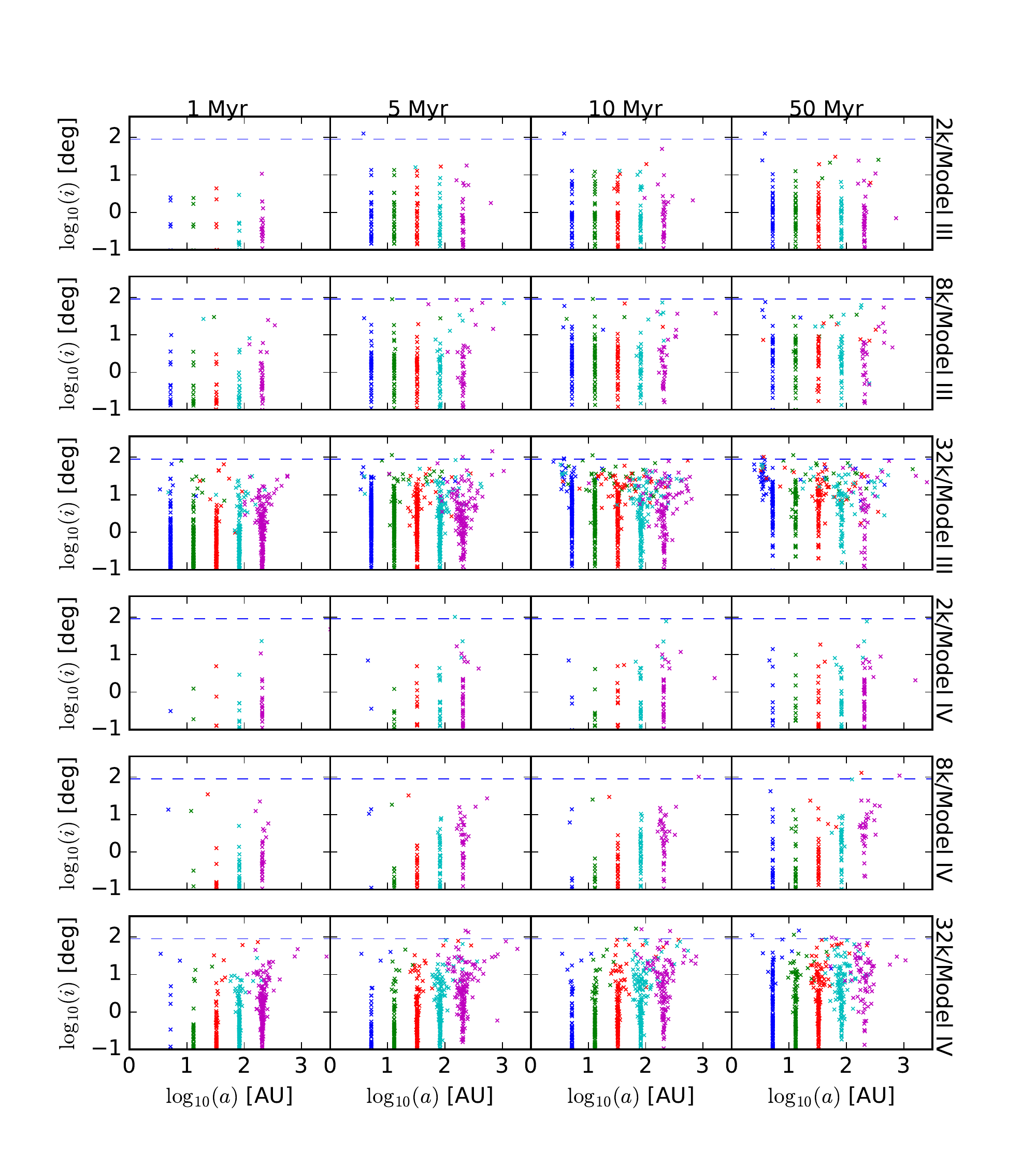}
\caption{Same with Figure~\ref{fig:ae_snapshots}, but for the $a-i$ space. Each panel is divided into two regimes by the blue horizontal dashed line: prograde orbits ($0^{\circ} \leq i < 90^{\circ}$) and retrograde orbits ($90^{\circ} \leq i < 180^{\circ}$). }
\label{fig:ai_snapshots}
\end{figure*}

All planets are initially coplanar, and external torques outside the orbital plane are exerted by perturbers from arbitrary directions. Mutual inclinations form as a natural byproduct of stellar encounters. Therefore, the excitations of orbital eccentricities are usually accompanied by the excitations of inclinations, which is consistent with the results from \cite{parker2012}. Figure~\ref{fig:ai_snapshots} shows snapshots a grid of $a-i$ space, where the inclinations are measured with respect to the initial orbital planes of planets. A small number of planets have been induced to retrograde orbits. The fraction of retrograde orbits seems to be slightly higher than the multiple-Earth models (\textsc{Model II} and \textsc{Model IV}): there are 4 retrograde planets ($0.4\%$) in the 32k \textsc{Model IV} ensemble, but only 2 ($0.2\%$) in the 32k \textsc{Model III} ensemble. Also, in the 8k cluster, the \textsc{Model IV} ensemble has two retrograde orbits, comparing to no retrograde orbits in the corresponding \textsc{Model III} ensemble. Given the low numbers of retrograde orbits, we are not sure yet whether this comparison is statistically significant. We believe that this result is consistent with theoretical understanding: planets in multiple-Earth systems are very weakly coupled ($k=100$), and carry much less orbital angular momentum than planets in the multiple-Jupiter systems. Consequently, it is easier to flip over the orbits in multiple-Jupiter systems.

%%%%%%%%%%%%%%%%%%%%%%%%%%%%%%%%%%%%%%%%%%%%%%%%%%%%%%%
%%%%%%%%%%%%%%%%%%%%%%%%%%%%%%%%%%%%%%%%%%%%%%%%%%%%%%%
%%%%%%%%%%%%%%%%%%%%%%%%%%%%%%%%%%%%%%%%%%%%%%%%%%%%%%%

\section{Conclusions} \label{sec:conclusions}

The collapse of giant molecular clouds triggers star formation in clustered environments. Protoplanetary disks, which are the progenitors of planets, form around newborn stars as byproducts of this process. Both theoretical predictions and observation suggest that planets are common, but only very few exoplanets have been discovered in star clusters. To better understand this apparent dichotomy, we carry out this exploratory study with a grid of simulations to test the dynamical stability of multiplanetary systems in intermediate-mass open clusters. We simulate three host star cluster environments with $N=2\kilo, 8\kilo$ and $32\kilo$ Plummer models. Each of these models has the same virial radius $R_v = 1$~pc. The mass spectra of these clusters are sampled with a \cite{kroupa2001} initial mass function. For each star cluster model, we distribute an ensemble of an ensemble of equal-mass and equally separated (in terms of mutual Hill radii) planetary systems around solar-type stars. Each planetary system has either five $\mjup$ (multiple-Jupiter model) or five $M_{\oplus}$ (multiple-Earth model) planets distributed in initially circular and coplanar orbits. The star clusters are integrated using the direct $N$-body code \nbodylong{} \citep{wang2015a, wang2016, spurzem99}, while the planetary systems are evolved with \rebound{} \citep{rein12} using the IAS15 integrator \citep{rein15}. The star cluster data is stored in the Block Time Step storage scheme \citep{cai2015}. After performing interpolation on the GPU, the perturbation information is passed to \texttt{rebound} within the \amuse{} \citep[][]{portegies09, portegies13, mcmillan2012, Pelupessy13} framework. Our conclusions can be summarized as follows.

\begin{enumerate}

	\item We quantify the strength of each stellar encounter with the dimensionless parameters $K$ and $V_{\rm inf}$, where $K$ is essentially the ratio between the perturbation timescale and the orbital timescale, and $V_{\rm inf}$ is the speed of the perturber at infinity. The peak frequency distribution of $K$ shifts to a lower values in denser clusters, indicating that the encounters are on average stronger in denser clusters. Moreover, stellar encounters are more frequent in denser and more massive clusters.

	\item The dynamical evolution of planetary systems is sensitive to external perturbations. Consequently, the planet survival rate declines in denser clusters: for clusters with $N=2\kilo, 8\kilo$ and $32\kilo$, the planet survival rates of the compact multiple-Jupiter systems (\textsc{Model I}) are $98.4\%, 94\%$ and $83.4\%$, respectively, and the survival rates for wide multiple-Jupiter systems (\textsc{Model III}) are $92\%, 79.2\%$ and $55\%$, respectively. Similarly, when evolving the compact multiple-Earth systems (\textsc{Model II}) in the $N=2\kilo, 8\kilo$ and $32\kilo$ clusters, the corresponding survival rates are $98.8\%, 98.6\%$ and $91.6\%$, respectively, and for wide multiple-Earth systems (\textsc{Model IV}) $96.8\%, 90\%$ and $77.1\%$, respectively.  In terms of the number of planets survived in each planetary system, $84\%$ of wide multiple-Jupiter systems and $90\%$ of wide multiple-Earth systems are able to keep all their planets $N=2\kilo$ cluster. This fraction drops to $26.5\%$ and $43\%$ in the $N=32\kilo$ denser cluster. Therefore, we believe that young low-mass star clusters will be prominent sites for next generation planet detection surveys, but the likelihood of detecting planets in dense globular clusters such as 47 Tuc would be low.

	\item External perturbations constrain the maximum sizes of planetary systems. The wider a planet's orbit is, the more vulnerable it is to external perturbations. In all star clusters environments used in our simulations, the survival rates of the wide models (\textsc{Model III} and \textsc{Model IV}) are much lower than the corresponding compact models (\textsc{Model I} and \textsc{Model II}), even though they evolve in exactly the same host star in the same cluster. As such, we predict that planets in denser star clusters will have smaller orbits, which actually allow them to be detected relatively easily.

	\item Planet-planet interactions are the catalysts of planetary system disruptions. We compare the dynamical evolution of the multiple-Jupiter systems and multiple-Earth systems in identical stellar environments, and found that the multiple-Earth systems in the all clusters have substantially  higher survival rates than the multiple-Jupiter systems. In multiple-Earth systems, inner planets absorb the angular momentum deficit of the outer planets through mutual interactions, and consequently leads to transfer of eccentricities from outer planets to inner ones. In multiple-Earth systems, planet mutual interactions are negligible, hence the eccentricity excitations of each planet is solely induced by external perturbations. 
	
	\item The excitation process is cumulative and gradually results in planet ejections. While very strong encounters can cause instantaneous ejection of planets, they are rare: only about $3\%$ of the encounters are strong enough to cause an eccentricity excitation of $\Delta e \geq 0.5$. In most cases planets are excited gradually by a series of moderate or weak encounters. 
	
	\item One direct consequence of stellar encounters is that they change both the magnitudes and the directions of orbital angular momenta. The changes in angular momenta are primarily caused by perturbers with offsets of the orbital planets, which is usually the case. In the EMS model for planetary systems we distribute planets initially on coplanar circular orbits. As such, the orbital angular momenta are proportional to $\sqrt{a}$, where $a$ is the semi-major axis. However, outer planets lose their angular momenta quickly though the rapid eccentricity excitations, and therefore they tend to have a higher inclinations than the inner planets. We observe on average $1-2$ planets ($0.1\% - 0.2\%$) with retrograde orbits in each multiple-Jupiter ensemble, and $2-4$ in the multiple-Earth ensembles. The frequency of planets with retrograde orbits increases in denser stellar environments. When planets with highly eccentric orbits approach the host stars, their eccentricities could be damped near the periapsis through the tidal damping effect, resulting to very small orbits (e.g. hot Jupiters). The tidal damping effect does not affect their induced inclinations. We speculate that this may produce an alternative channel of producing large spin-orbit misalignments (as opposed to the Kozai-Lidov mechanism), which is observed among many extrasolar planetary systems through e.g., the Rossiter-McLaughlin effect.
	
%	\item Short period planets can be produced by a combination of stellar encounters and planet-planet scattering. Depending on the stellar density, there are roughly $5 \%$ of planets with $a \leq 1.0$ AU and $r_{\rm p} \leq 0.1$ AU, where $a$ and $r_{\rm p}$ are the semi-major axis and the periapsis distance of the orbit. These planets are candidates of Hot Jupiters, if the tidal circularization process is effective. Outer planets are ejected more easily, and once ejected, the remaining planetary system may actually be more stable, as long as the remaining planets are protected in the potential well of the host star, their orbits are sparely populated and/or tidal circularization are operating on them. {\bf data to be updated. Make a table?}
		
\end{enumerate}

Not all star clusters are as compact as our models. As such, our results provides an upper limit of planet ejection rates in such environments.

It is worthwhile to point out that the background cluster potential may have implications to planetary system stability. For an isolated planetary system, the perturber and the host stars are interacting in the Keplerian potential, and the total energy is conserved. In the star cluster environments (especially in the cluster center), however, stars are being scattered randomly, exhibiting Brownian motion in the background cluster potential. Therefore, our simulations of perturbing planetary systems in star clusters cannot be simplified as isolated planetary systems being perturbed by a series of Keplerian stellar encounters. The background cluster potential affects the planet stability by affecting the trajectories of the perturber. Nevertheless, a quantitative analysis of this effect is beyond the scope of this paper.

This study has been limited to pure dynamical interaction between member stars in star clusters and multiplanetary systems. Planets are assumed to be of equal mass, arranged in initially coplanar circular orbits with equal separation in terms of their mutual Hill radii. Eccentricity damping due to the protoplanetary disks and/or tidal circularization process may contribute to the robustness of planetary systems. In addition, our host star clusters are sampled with the idealized Plummer model in virial equilibrium ($Q=0.5$). In reality, planetary systems are immensely diverse in terms of orbital architectures and mass spectra. Host star clusters may depart from $Q=0.5$ and exhibit substructures and \citep{zheng2015,portegies2016}. It is also possible that the chaotic behavior observed in some mildly perturbed planetary systems only manifests itself when simulating for a more extended time. 

Nevertheless, this study aims to highlight the dynamical consequences of stellar encounters for planetary systems in star clusters. The results are also potentially insightful for understanding the frequency of free-floating planets, which are currently subjected to strong observational bias due to the difficulty of observations. 

\section*{Acknowledgements}
We thank the anonymous referee for her/his comments that helped to improve the manuscript considerably. We thank Adrian Hamers for commenting on the manuscript. We thank Douglas N.C. Lin, Fred C. Adams, Sverre Aarseth, Roberto Capuzzo-Dolcetta, Rosemary Mardling, and Beibei Liu for useful discussions. We thank Long Wang for graciously providing support for the direct $N$-body code \nbodylong{}.
We are grateful for the support from the \amuse{} team, in particular Inti Pelupessy, Arjen van Elteren, Nathan de Vries, Michiko Fujii and Lucie J{\'{\i}}lkov{\'a} for very helpful  discussions. MBNK acknowledges the AMUSE team for supporting his visit to Leiden University.
We are grateful for support by Sonderforschungsbereich SFB 881 ``The Milky Way System'' of the German Research Foundation (DFG), through subproject Z2 and the GPU cluster Milky Way at FZ J\"ulich, and for the support of the visit of MXC in Heidelberg.
We acknowledge support by NAOC CAS through the Silk Road Project and (RS) through the Chinese Academy of Sciences Visiting Professorship for Senior International Scientists, Grant Number $2009S1-5$. The special GPU accelerated supercomputer {\tt laohu} at the Center of Information and Computing at National Astronomical Observatories, Chinese Academy of Sciences, funded by Ministry of Finance of People's Republic of China under the grant $ZDYZ2008-2$, has been used for some of the largest simulations.
M.B.N.K. was supported by the National Natural Science Foundation of China (grants 11010237, 11050110414, 11173004, and 11573004). This research was supported by the Research Development Fund (grant RDF-16-01-16) of Xi'an Jiaotong-Liverpool University (XJTLU).

This work was supported by the Netherlands Research Council NWO (grants \#643.200.503, \#639.073.803 and \#614.061.608) by the Netherlands Research School for Astronomy (NOVA). This research was supported by the Interuniversity Attraction Poles Programme (initiated by the Belgian Science Policy Office, IAP P7/08 CHARM) and by the European Union's Horizon 2020 research and innovation programme under grant agreement No 671564 (COMPAT project).

%%%%%%%%%%%%%%%%%%%%%%%%%%%%%%%%%%%%%%%%%%%%%%%%%%



%\bibliographystyle{mnras}


%%%%%%%%%%%%%%%%%%%%%%%%%%%%%%%%%%%%%%%%%%%%%%%%%%

%%%%%%%%%%%%%%%%% APPENDICES %%%%%%%%%%%%%%%%%%%%%

%\appendix

%%%%%%%%%%%%%%%%%%%%%%%%%%%%%%%%%%%%%%%%%%%%%%%%%%

% Don't change these lines
\bsp	% typesetting comment
\label{lastpage}
\end{document}